\documentclass[structabstract]{aa}
\usepackage{graphicx}
\usepackage{txfonts}
\usepackage{graphicx}
\usepackage{natbib}
\usepackage{epsf,subfigure}
\usepackage{amssymb}
\newcommand{\D}{\mathrm{d}}
\newcommand{\I}{\mathrm{i}}
\newcommand{\sgn}{\mathrm{sgn}}
\begin{document}
   \title{Elliptical instability in terrestrial planets and moons}

   \subtitle{}

   \author{D. Cebron
          \inst{1}\fnmsep\thanks{corresponding author: cebron@irphe.univ-mrs.fr}
          \and
          M. Le Bars\inst{1}
          \and
          C. Moutou\inst{2}
          \and
          P. Le Gal\inst{1}
          }

   \institute{Institut de Recherche sur les Ph\'enom\`enes Hors Equilibre, UMR 6594, CNRS and Aix-Marseille Universit\'e, 49 rue F. Joliot-Curie, B.P. 146, 13384 Marseille cedex 13, France.
         \and
             Observatoire Astronomique de Marseille-Provence, Laboratoire d'Astrophysique de Marseille, 38 rue F. Joliot-Curie, 13388 Marseille cedex 13, France.
             }

  % \date{Received ?, 2011; accepted ?, 2011}

% \abstract{}{}{}{}{}
% 5 {} token are mandatory

  \abstract
  % context heading (optional)
  % {} leave it empty if necessary
   {The presence of celestial companions means that any planet may be subject to three kinds of harmonic mechanical forcing:   tides, precession/nutation, and libration.
   These forcings can generate flows in internal fluid layers, such as fluid cores and subsurface oceans, whose dynamics then significantly
   differ from solid body rotation. In particular, tides in non-synchronized bodies and
   libration in synchronized ones are known to be capable of exciting the so-called elliptical instability,
   i.e. a generic instability corresponding to the destabilization of two-dimensional flows with elliptical streamlines,
   leading to three-dimensional turbulence. }
  % aims heading (mandatory)
   {We aim here at confirming the relevance of such an elliptical instability in terrestrial bodies
  by determining its growth rate, as well as its consequences on energy dissipation,
   on magnetic field induction, and on heat flux fluctuations on planetary scales.}
  % methods heading (mandatory)
   {Previous studies and theoretical results for the elliptical instability are re-evaluated and extended to cope with an astrophysical context. In particular, generic analytical expressions of the elliptical instability growth rate are obtained using a local WKB approach,
   simultaneously considering for the first time (i) a local temperature gradient due to an imposed temperature contrast across the considered layer or to the presence of a volumic heat source and
   (ii) an imposed magnetic field along the rotation axis, coming from an external source.}
  % results heading (mandatory)
   {The theoretical results are applied to the telluric planets and moons of the solar system as well as to three Super-Earths:
   55 CnC e, CoRoT-7b, and GJ 1214b.
   {For the tide-driven elliptical instability in non-synchronized bodies, only the Early Earth core is shown to be clearly unstable.
   For the libration-driven elliptical instability in synchronized bodies,
   the core of Io is shown to be stable, contrary to previously thoughts, whereas Europa, 55 CnC e, CoRoT-7b and GJ 1214b cores can be unstable. The subsurface ocean of Europa is slightly unstable}. However, these present states do not preclude more unstable situations in the past.}
  % conclusions heading (optional), leave it empty if necessary
   {}

%    \keywords{Terrestrial planets --
%         Elliptical instability  --
%         Librations -- Tides --
%                 WKB analysis
%                }

   \maketitle
%
%________________________________________________________________

\section{Introduction}

The flows in fluid layers of planets and moons are of major interest because they imply first order consequences for their internal dynamics and orbital evolutions. Indeed, internal flows
create torques on solid layers and induce energy dissipation. Moreover, internal flows are directly
responsible for the generation of magnetic fields, either by induction of an
existing background magnetic field or by excitation of a
self-sustained dynamo. Finally, planetary heat fluxes are also
directly linked to flows in fluid layers, which can act as thermal
blankets for stably stratified configurations, or as efficient heat
flux conveyers in the case of convective flows.

Planetary fluid layers are subject to body rotation, which implies
that inertial waves can propagate through them
\cite[e.g.][]{greenspan1968}. Usually damped by viscosity, these
waves can, however, be excited by longitudinal libration, precession, and tides,
which are harmonic mechanical forcings of azimuthal periodicity
$m=0$, $1$, and $2$, respectively. The fluid response to such forcings
in ellipsoids is a long-standing issue: see e.g. for longitudinal libration
\cite{aldridge1969axisymmetric}, \cite{noir2009experimental}, \cite{calkins2010axisymmetric}, \cite{sauret2010experimental}, \cite{chan2011simulations} and \cite{zhang2011fluid}, for latitudinal libration, \cite{chan2011simulations-2}, for
precession, \cite{Poincare_precession}, \cite{busse1968steady}, \cite{cebron2010tilt}, \cite{kida2010super,kida2011steady} and \cite{zhang2010fluid}, and for tides, \cite{ogilvie2004tidal}, \cite{ogilvie2007tidal}, \cite{tilgner2007zonal}, \cite{rieutord2010viscous} and \cite{morize2010experimental}. In these studies, it has been shown that the dynamics of
a fluid layer is completely modified when the forcing resonates with
an inertial wave. In addition to these direct forcings, inertial
waves can also form triadic resonances, leading to parametric
inertial instabilities. For instance, the so-called shear
instability can be excited by precession
\cite[][]{kerswell1993instability,lorenzani2001fluid,lorenzani2003inertial},
and the elliptical instability can be excited by tides in
non-synchronized bodies
\cite[][]{malkus1989experimental,rieutord2000note} and by librations in
synchronized ones \cite[][]{kerswell1998tidal}.

The elliptical instability is a generic instability that affects any
rotating fluid whose streamlines are elliptically deformed \cite[see
the review by][]{kerswell2002elliptical}. A fully three-dimensional
turbulent flow is excited in the bulk as soon as (i) the ratio
between the ellipticity of the streamlines $\beta$ and the square
root of the Ekman number $E$ (which represents the ratio between the
viscous over the Coriolis forces) is more than a critical value on the order of one and (ii) as soon as a difference in angular velocity
exists between the mean rotation rate of the fluid and the
elliptical distortion. In a planetary context, the ellipticity of
streamlines is related to the gravitational deformation of all
layers of the considered body, coming from the static and periodic
terms of the tidal potential, as well as from a potential frozen
bulge. The differential rotation between the fluid and the
elliptical distortion can be oscillatory when caused by libration in
synchronized systems, or stationary in non-synchronized ones. The
elliptical instability is then respectively refered to as libration-driven
elliptical instability (LDEI) and tide-driven elliptical
instability (TDEI). TDEI and LDEI have already been
suggested as taking place respectively on Earth
\cite[e.g.][]{aldridge1997elliptical} and on Io
\cite[e.g.][]{kerswell1998tidal}. However, these previous works do
not consider some planetary particularities, so they need to be
revisited. For instance, \cite{aldridge1997elliptical} did not take the orbital rate of the Moon into account or the magnetic field of
the Earth, thus neglecting the effects of tide rotation and Joule
dissipation on the growth of TDEI. \cite{kerswell1998tidal}
implicitly assumed that the tidal response of Io is completely
fluid, neglecting the rigidity of its mantle and overestimating the
amplitude of librations and tidal deformations. Our purpose here is
to extend previous results of the literature on TDEI and LDEI and
to determine general formulas for quantifying the presence of the
elliptical instability in terrestrial bodies, taking the relevant complexities present in natural systems into account.

This paper is organized as follows. Section \ref{sec:forcing}
presents the different celestial forcings that could excite an
elliptical instability, first focusing on tides in non-synchronized
systems, and then on forced and free libration in synchronized ones.
In section \ref{sec:context}, we introduce our physical model and
develop a local WKB analysis in all configurations, including the
effects of viscosity, as well as the effects of an imposed magnetic
field and a local temperature gradient. These theoretical results
are used in section \ref{sec:applications} to investigate the possible
presence of any of TDEI and LDEI in telluric planets and moons of
the solar system, as well as in two Super-Earths of extrasolar
systems. The possible consequences of those instabilities are
finally considered.

\section{From celestial mechanics to the excitation of an elliptical instability} \label{sec:forcing}

\begin{figure}                  % Chaque figure doit avoir pour nom nomfig1.eps,
                                        % ou Nom est celui du premier auteur nomfig2.eps
  \begin{center}
%     \begin{tabular}{ccc}
      \setlength{\epsfysize}{6.0cm}
      \subfigure[]{\epsfbox{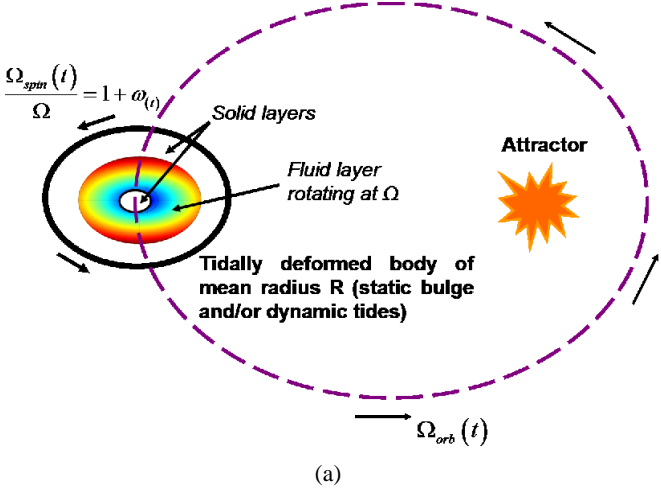}} \\

    \caption{Sketch of the problem studied in this work. With the mean rotation rate of the fluid $\Omega$,
    we define the dimensionless orbital rotation rate $\gamma_{(t)}$ and the dimensionless spin rotation rate $1+\omega_{(t)}$. The phase lag between the tide and the gravitational potential of the host body is not relevant to our purpose.}
    \label{cebronfig01}             % Pensez ?? mettre le nom du premier auteur ?? la place de Nom
  \end{center}
\end{figure}

Figure \ref{cebronfig01} presents a sketch of the problem considered
in this work. We consider a telluric body of rotation rate
$\Omega_{spin}$, orbiting around an attractor (orbit in black dashed
line) at the orbital rate $\Omega_{orb}$. This body has a radius
$R$, a mass $M$, and a fluid layer in its interior between an
external radius $R_2$ and an internal radius $R_1$, typically a
liquid outer core. We suppose that this internal fluid layer is
enclosed between an external elliptically deformed solid layer and a
possible inner core, such as the outer liquid core of the Earth or
the subsurface ocean of Europa. The elliptical deformation
can admit different origins. First, in the presence of an orbiting
companion, the elliptical deformation can come from the static and
periodic terms of the tidal potential as seen from the mantle frame
of reference. In this case, periodic terms lead to tides, and the
static term leads to the so-called static (tidal) bulge (the solid layer behaves as a fluid layer on the long
term). Second, a so-called frozen bulge, resulting from previous states, may exist, as for
instance in the Moon \cite[][]{garrick2006evidence}. In this case, the body is not in
hydrostatic equilibrium. For our purpose, it is sufficient to
distinguish between the response to tidal potential periodic terms,
which we call the dynamic tides, and a permanent (or very slowly
changing) bulge, either owing to the tidal potential static terms or
due to a frozen bulge, which we call a static bulge. The
usual phase lag between the tide and the gravitational potential of
the host body, which is due to internal dissipation, is not relevant
for our purpose so is forgotten. Three dimensionless numbers
are needed to describe the system: (i) the ellipticity $\beta$ of
the elliptical deformation, (ii) the Ekman number $E=\nu/(\Omega\
R_2^2)$, where $\nu$ is the fluid kinematic viscosity, $R_2$ the
outer radius of the rotating fluid, and $\Omega$ its typical angular
velocity before any instability, equal to the mean value of the
(possibly varying) mantle spin rate $\Omega_{spin}(t)$, (iii) the
differential rotation $\triangle \Omega $ between the fluid and the
elliptical distortion, non-dimensionalized by the fluid rotation
rate, $\bigtriangleup \Omega / \Omega$. We distinguish two cases: a non-synchronized body, and a synchronized
body. In the former, over one spin period, a mean differential
rotation exists between the elliptical deformation and the fluid,
whereas in the latter the mean rotation rates of the
deformation and of the fluid are equal. The different cases are
described in the following and summarized in Table
\ref{table_intro}.

\begin{table}
\caption{List of the different astrophysical configurations that
could lead to an elliptical instability (E.I) in a planetary fluid layer
(liquid core, subsurface ocean) of a non synchronized or a synchronized celestial body.} \label{table_intro}

\begin{tabular}{@{}l|c|c|c|c|c}
State       & Origin& Origin
& $\bigtriangleup \Omega$  & E.I   \\
       &of $\bigtriangleup \Omega$ &of $\beta$
&  &    \\
\hline
Non-sync.      & spin rotation & D.T.${}^a$ & $\Omega_{spin}-\Omega_{orb}$  & TDEI  \\
Non-sync.      &  impact    & S.B.${}^a$     & spin-up process   & TDEI   \\
Sync.  &  forced O.L${}^b$   &  D.T. & $2e\Omega \cos(2 \pi t/T_{orb})$     & LDEI     \\
Sync.  &  forced P.L${}^b$   &  S.B.    & $\epsilon \Omega \cos(2 \pi t/T_{orb})$  & LDEI   \\
Sync.  &  free P.L    &  S.B.  & $\epsilon \Omega \cos(\omega_{free} t)$    & LDEI  \\
Sync.  &  any P.L    &  S.B., D.T.   & zonal flow${}^c$  & TDEI  \\
 \end{tabular}
\bigskip
$\beta$ is the ellipticity of the boundaries distortion, $\bigtriangleup \Omega$ is the differential rotation rate between the fluid and the elliptical deformation, $\Omega$ is the mean spin rate of the planet, $\epsilon$ is the physical libration amplitude, $T_{orb}$ and $e$ are the orbital period and eccentricity, and $\omega_{free}$ is the free libration frequency.\\

${}^a$ D.T. stands for dynamic tides and S.B stands for static bulge.\\
${}^b$ O.L. and P.L stand respectively for optical and physical libration.\\
${}^c$ Case equivalent to a non-synchronized case (cf. section \ref{sec:free}).
\end{table}

\subsection{Non-synchronized bodies}

For a non-synchronized body, we consider two cases depending on the
origin of the elliptical shape. First, if the spin rate of the mantle $\Omega_{spin}$ is constant, a
TDEI can be excited by the tidal elliptical distortion due to dynamic
tides, which rotate at the orbital velocity $\Omega_{orb}$. This is the standard
configuration considered for instance by \cite{craik1989stability}, who showed
that TDEI is indeed possible except in a forbidden zone
$\Omega_{spin}/\Omega_{orb} \in [-1;1/3]$, where no triadic
resonance is possible.

Second, if the elliptical shape comes from a static bulge, a non-zero
mean differential rotation over one spin period implies that the
fluid does not rotate at the same rate as the mantle, which
corresponds to a spin-up or a spin-down process. This can occur, for
instance, transiently after a large meteoritic impact, which is capable of
fully desynchronizing the body \cite[see for instance the considered
scenario for explaining the Moon's magnetic field by][]{lebarsNature}. In this case, a differential rotation exists between the
fluid and the mantle with its static bulge, up to the typical
spin-up/spin-down time necessary for the fluid to recover the mantle
velocity  \cite[][]{greenspan1968}, i.e. up to
\begin{eqnarray}
t_{spin-up}=\Omega_{spin}^{-1}\ E^{-1/2}. \label{spinup}
\end{eqnarray}
Then, if the growth time of the TDEI is short
enough compared to the spin-up time, one can expect a quasi-static
evolution of the system, where the modification of the spin rate of
the fluid is neglected during the growth of the instability: the
former configuration is then transiently recovered.

\subsection{Synchronized bodies}
In the synchronized case, even if there is no mean
differential rotation between the elliptical deformation and the
fluid, oscillations can nevertheless occur for different reasons.
For the study of the elliptical instability, it is necessary to know
the amplitude of these oscillations, which depends on their origins.
We distinguish below the forced librations caused by gravitational
interactions with other celestial bodies, and free librations
induced for instance by a meteoritic impact.

\subsubsection{Forced librations} \label{sec:forced_lib}

In forced librations, static bulge and dynamic tides have to be
considered simultaneously. To illustrate this, following
\cite{goldreich2010elastic}, we consider a simple toy model without
any internal dissipation: a synchronously spinning satellite, with
an elastic outer shell and a homogeneous fluid interior, moving on
an elliptic orbit. The orbital velocity changes along the orbit, and
writes at first order in the orbital eccentricity $e$ as
\begin{eqnarray}
\Omega_{orb}=\Omega\ (1+2e \cos{\Omega t}),
 \label{eq:orbit}
\end{eqnarray}
where $\Omega$ is the mean value of the mantle spin rate
$\Omega_{spin}(t)$. Considering the influence of the orbital velocity variations on the
satellite dynamics, we expect two limit cases: (i) if the rigidity
of the elastic shell is zero or if the planet spin rate is
low enough for the shape of the planet to have time to adapt to
the gravitational constraints, the shell slides over the fluid and
maintains its equilibrium shape, with the long axis of the
ellipsoidal figure pointing toward the companion body; (ii) if the
rigidity of the elastic shell is strong enough or if the
planet spin rate is rather high, the entire satellite rotates
rigidly with a fixed shape.

In the first case, only the
elastic energy $E_{elas}$ varies: the meridians of the shell are
stretched and compressed due to the rotation, whereas the spin
velocity of the satellite remains constant. This is the so-called
optical libration.  In the second case, only the gravitational energy
$E_{grav}$ varies and the spin velocity of the satellite changes, which corresponds to the so-called physical libration.

In both cases, the libration period remains small compared
to the typical spin-up/spin-down time (\ref{spinup}), which means
that the fluid does not follows the solid boundaries because it never has
enough time to adapt to the periodic velocity fluctuations and
continues to rotate at the constant synchronous rotation rate
$\Omega$. This is the so-called no spin-up condition. In
the first case, therefore, a differential rotation exists between the fluid
rotating at the constant rate $\Omega$ and the dynamic tides
rotating at the oscillating orbital velocity, $\bigtriangleup \Omega
/ \Omega = 2e\cos{\Omega t}$. An LDEI can thus be excited by this
optical libration, as shown theoretically by
\cite{kerswell1998tidal} and \cite{herreman2009effects}. In the
second, a differential rotation $\bigtriangleup \Omega / \Omega
= \epsilon \cos{\Omega t}$ exists between the fluid rotating at the
constant rate $\Omega$ and the static bulge subject to physical
librations of amplitude $\epsilon$, which depends on the internal
structure of the satellite. The amplitude of the
physical librations $\epsilon$ is always less than the $2e$
extreme value given by optical libration, because of different
internal torques such as the gravitational torque and elastic strain torque \cite[see for instance][]{van2008librations,van2009effect}.

With the more general case of an arbitrary torque applied to the
shell, \cite{goldreich2010elastic}  estimated the ratio $\Re =
E_{elas}/E_{grav}$ by
\begin{eqnarray}
\Re = \frac{32\ \pi}{5}\ \frac{1+\tilde{\nu}}{5+\tilde{\nu}}\
\frac{(1+k_f)^2}{k_f}\ \frac{\tilde{\mu} d R^3}{GM^2}
\end{eqnarray}
where $\tilde{\nu}$ is the Poisson ratio, $k_f$ the fluid Love
number, $\tilde{\mu}$ the shell rigidity, $R$ and $d$ the mean
radius and the thickness of the shell, $M$ the mass of the
satellite. According to \cite{goldreich2010elastic}, typical values
give $\Re \sim 10^{-2}$ for the subsurface ocean of Europa, and $\Re
\sim 0.1$ for the subsurface ocean of Titan, whereas the silicate
mantle of Io is expected to behave in the limit $\Re \gg 1$. Because
of the visco-elastic rheology of real bodies, the effective response
should be between these two extrem cases given by this model.
\cite{goldreich2010elastic} argue that the total increase in energy
would be minimal, which leads us to consider that Europa and Titan,
for instance, behave like entirely fluid satellites. In contrast, \cite{karatekin2008effect},
\cite{van2008librations,van2009effect}, and
\cite{baland2010librations} consider that the arguments
proposed by \cite{goldreich2010elastic} are unrealistic on short
timescales, hence that the rheology does not allow the bodies to
reach their minimal energetic state. They assume that they behave
rigidly, however, large libration may still be due
to resonances with free libration modes, which may be reached for
thin ice shells \cite[][]{baland2010librations}. In
addition to the displacements of the ice shell and mantle induced by
gravitational interactions, relatively large longitudinal
displacements may also be induced directly in the fluid layers by
the periodical part of the tidal potential \cite[][]{tobie2005tidal}.
This would constitute a supplementary origin for a non-zero
differential rotation between the fluid and the deformation. Either way,
all these issues are still being debated and are clearly beyond the scope
of the present paper. All that is needed here is to know the
amplitude of the relative motion between the elliptical distortion
and the fluid. In the following, we consider the full range of
configurations up to a maximum distortion amplitude given by the
extreme value of optical libration.

\subsubsection{Free librations} \label{sec:free}

After a meteoritic impact, for instance, so-called free librations
can occur on the typical resynchronization time
\cite[e.g.][]{williams2001lunar}. Following the no spin-up condition
explained in section \ref{sec:forced_lib}, the fluid keeps rotating
at the orbital velocity (synchronized state), while the mantle
librates around this mean value. The amplitude $\epsilon$ of the
free librations depends initially on the impact strength and
decreases through time, and the libration frequency remains
equal to a proper frequency of the body, given by $\omega_{free}=
\Omega\sqrt{3 (B-A)/C}$ at first order in the orbital eccentricity, where $(A,B,C)$ are the three principal moments of inertia of the body
\cite[see for instance][]{lissauer1985can}. Considering a static
bulge, free librations can thus drive an LDEI from the differential
rotation $\bigtriangleup \Omega / \Omega = \epsilon
\cos({\omega_{free} t})$, providing that the growth time of the
instability is shorter than the resynchronization time, as shown in section \ref{sec:orblib}.

\subsubsection{Zonal wind induced by physical librations}

Finally, in all scenarios involving physical librations, it has
recently been determined analytically by \cite{busse2010mean}, and
confirmed experimentally and numerically by
\cite{sauret2010experimental}, that non-linearities in the Ekman
layer driven by the librating rigid boundaries induce a differential
rotation in the fluid of amplitude $\triangle \Omega/\Omega=-0.154\
(\theta \omega_o/\Omega)^2$, where $\theta$ is the amplitude angle
of the libration and $\omega_o$ its frequency. TDEI can thus be
excited by this differential rotation with both static bulge and
dynamic tides. Nevertheless, the differential rotation generated by this process is always very
small. We do not expect this mechanism to play an important role in
a planetary context, since it is always dominated by LDEI, but it is
worth here mentioning its existence since it may be relevant in
certain astrophysical cases.

\subsection{Typical amplitudes of gravitational distortions}

The amplitude $\beta$ of gravitational distortions, defined here as
$\beta=|a_{1}^2-a_{2}^2|/(a_{1}^2+a_{2}^2)$, where $a_1$ and $a_2$
are respectively the long and short axes of the outer boundary of
the considered fluid layer, is generally unknown for celestial
bodies. To study the elliptical instability for real cases, we need to estimate it, for instance
by assuming an hydrostatic equilibrium shape.

The equilibrium shape of a body of mass $M$ and radius $R$, is an
old problem that begins with the static bulge theory of
\cite{newton1999philosophiae}. This classical theory considers an
incompressible no-spinning body at rest, deformed by a tidal field at leading order in $R/D$, which leads to a spheroidal shape and
\begin{eqnarray}
\beta=\frac{3}{2} \frac{M_2}{M}\frac{R^3}{D^3}, \label{eq:newton}
\end{eqnarray}
where $M_2$ is the mass of the body responsible for the
gravitational field and $D$ the distance between the two bodies.
This tide is sometimes referred to as the marine tide, where the gravitational potential of the tidal bulge is neglected. This approximation always leads to a relevant but
underestimated tidal deformation. When possible, we use in the following a better estimate of $\beta$ that takes the density
distribution in the body and the gravitational potential of the tidal bulge into account:
\begin{eqnarray}
\beta=\frac{3}{2}\ h_2\ \frac{M_2}{M}\frac{R^3}{D^3} \label{eq:newton2}
\end{eqnarray}
with the radial displacement Love number $h_2$, directly linked to
the potential Love number $k_2$ by $h_2=1+k_2$. A typical value is
$k_2=3/2$, obtained for an incompressible homogeneous body in
hydrostatic equilibrium \cite[e.g.][]{greff2005analytical}. The
tidal Love numbers can be calculated with the Clairaut-Radau theory
\cite[see e.g.][]{van2008librations}.

As shown by equation (\ref{eq:newton2}), gravitational distortions
vary with the interbody distance $D$. They can thus be divided into
a component of constant amplitude, corresponding to the mean value
of the gravitational distortion along the elliptic orbit, plus a
smaller component with an amplitude oscillating between $\pm 3e$
times the constant one \cite[e.g.][]{greenberg2003tidal}. In real
cases that we consider in the following, this oscillating
component can be neglected since it will always have a second-order
influence on the elliptical instability compared to the constant
(static or dynamic) component of $\beta$ (but see appendix
\ref{sec:diurnal}). Nevertheless, in synchronized
satellites, these so-called diurnal tides have important
consequences for the internal state and the orbital evolution, since
the changing shape of the bulge generates time-varying stresses,
which generate heat by viscosity or friction. Besides, as
shown by \cite{tobie2005tidal}, these diurnal tides should also induce
longitudinal motions in the fluid layers, which constitute a
supplementary origin of a non-zero differential rotation between the
fluid and the deformation. As already mentioned above, in the
context of the present study, this additional component is fully
included in the amplitude of the considered physical libration
available to drive the elliptical instability.

\section{Generic formulas for the growth rate of the elliptical instability in a planetary context} \label{sec:context}

\subsection{Model, equations, and dimensionless parameters} \label{sec:model}

\begin{figure}                  % Chaque figure doit avoir pour nom nomfig1.eps,
                                        % ou Nom est celui du premier auteur nomfig2.eps
  \begin{center}
      \setlength{\epsfysize}{6.0cm}
      \subfigure[]{\epsfbox{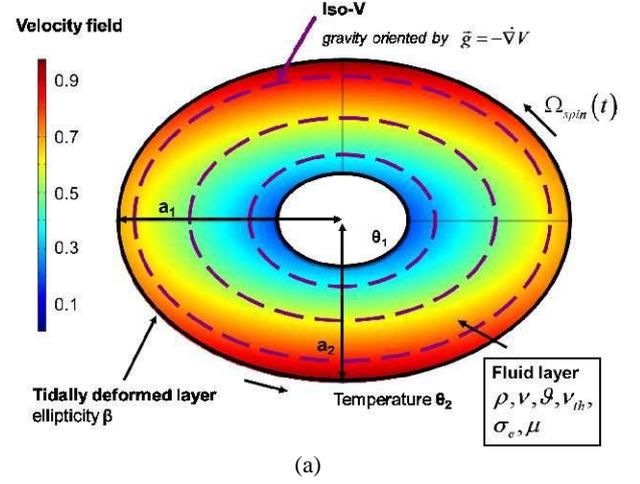}} \\
    \caption{Description of an internal liquid layer. The tidal forces deform the core-mantle boundary (CMB) in an ellipse of axes $a_1$ and $a_2$,
   which leads to an ellipticity $\beta$ of the streamlines. In this layer, the fluid is rotating at the rate $\Omega$.}
    \label{cebronfig1}             % Pensez ?? mettre le nom du premier auteur ?? la place de Nom
  \end{center}
\end{figure}

We consider a telluric celestial body in the general framework
sketched in figure \ref{cebronfig01}, and we focus on a liquid layer
described in figure \ref{cebronfig1}. All dimensional parameters are
listed in Table \ref{table_var}. The instantaneous spin rotation
rate $\Omega_{spin}=\Omega\ (1+\omega_{(t)})$ may depend on time
because of either free or forced physical librations. We focus on an
internal fluid layer enclosed in an ellipsoidal shell, with an outer
boundary of mean radius $R_2$ at temperature $\theta_2$, and an
inner boundary at temperature $\theta_1$, with a mean radius
$R_1=\eta R_2$. As already seen above, because of the no spin-up
condition, this fluid layer is initially rotating at the constant
rate $\Omega$, equal to the mean value of $\Omega_{spin}$. This
layer is considered to be homogeneous, with density $\rho_0$,
kinematic viscosity $\nu$, thermal expansion $\vartheta$, thermal
diffusivity $\nu_{th}$, electrical conductivity $\sigma_e$, and
magnetic permeability $\mu$. We focus here on the stability of the
elliptical flow in the equatorial plane, but note that our local
analysis remains valid in any plane orthogonal to the rotation axis.
We choose $R_2$ as the length scale and $\Omega^{-1}$ as the time
scale so that the mean basic spin of the body has an unit angular
velocity along the rotation axis $(O,\mathbf{e_{x_3}})$. The
elliptical deformation has a dimensionless angular velocity
$\gamma_{(t)}\ \mathbf{e_{x_3}}$, with $\gamma_{(t)}$ equal to
$\Omega_{orb}(t)/\Omega$ when looking at dynamic tides and to
$\Omega_{spin}(t)/\Omega$ when looking at static bulges (see Table
\ref{table_intro}).

We consider the frame where the elliptical distortion is fixed,
which is rotating at the angular velocity $\gamma_{(t)}
\mathbf{e_{x_3}} $, with $\mathbf{e_{x_1}}$ in the direction of the
long axis $a_1$ and $\mathbf{e_{x_2}}$ in the direction of the short
axis $a_2$. The dimensionless equations of fluid motions are
\begin{eqnarray}
\bigtriangledown \cdot \mathbf{u}=0, \label{eq:div}
\end{eqnarray}
\begin{eqnarray}
\frac{\partial \mathbf{u}}{\partial t}+2 \gamma\ \mathbf{e_{x_3}}
\times  \mathbf{u} + \frac{\D \gamma_{(t)}}{\D t}\ \mathbf{e_{x_3}}
\times  \mathbf{r} +  \mathbf{u} \cdot \mathbf{\nabla u} +
\mathbf{\nabla} p=E\ \nabla^2  \mathbf{u}+ \mathbf{f},
\label{Navier}
\end{eqnarray}
where $\mathbf{u}$ is the fluid velocity; $p$ the pressure
(including the centrifugal term) non-dimensionalized by $\rho_0\
{R_2}^2\ \Omega^2$; $E=\nu/(\Omega {R_2}^2)$ the Ekman number based
on the external radius; and $\mathbf{f}=\mathbf{f_B}+\mathbf{f_L}$
the volumic force, including the buoyancy force $\mathbf{f_B}$ and
the magnetic Lorentz force $\mathbf{f_L}$. The flow is rotating
within an ellipse $x_1^2/a_{1}^2+x_2^2/a_{2}^2=1$, and we define the
ellipticity as $\beta=|a_{1}^2-a_{2}^2|/(a_{1}^2+a_{2}^2)$.

\begin{table}
 \caption{List of the dimensional variables used in this work.}\label{table_var}
 \begin{tabular}{@{}lll}

 Orbital rotation rate &  $\Omega_{orb} (t)$  \\
 Spin rotation rate &  $\Omega_{spin} (t)$  \\
 Fluid rotation rate$^*$ &  $\Omega$  \\
 Mass of the deformed body & $M$\\
 Mass of the attractor & $M_2$\\
 Inter-body distance & $D$\\
 Free libration angular frequency & $\omega_{free}$ \\
 Fluid layer mean external radius & $R_2$\\   %$=\sqrt{(a_1^2+a_2^2)/2}$ \\
 Long/short axis in the equatorial plane & $a_{1}/a_{2}$ \\
 Fluid layer mean internal radius & $R_1=\eta R_2$ \\
 Imposed external temperature & $\theta_2$ \\
 Imposed internal temperature & $\theta_1$ \\
 Gravity at the external radius & $g_0$ \\
 Imposed magnetic field & $B_0\ \mathbf{e_{x_3}}$ \\
 Fluid density & $\rho_0$\\
 Fluid kinematic viscosity & $\nu$\\
 Fluid thermal expansion & $\vartheta$\\
 Fluid thermal diffusivity & $\nu_{th}$\\
 Fluid electrical conductivity & $\sigma_e$\\
 Fluid magnetic permeability & $\mu$\\
 \end{tabular}
\medskip

$^*$mean value of the spin rate.
\end{table}

Using the dimensionless temperature
$\theta=(\tilde{\theta}-\theta_2)/(\theta_1-\theta_2)$, the
temperature equation is
\begin{eqnarray}
\frac{\partial \theta}{\partial t} +(\mathbf{u} \cdot \nabla)\
\theta &=& \frac{E}{Pr}\ \bigl(\nabla^2 \theta -K\bigl),
\label{eq:theta}
\end{eqnarray}
where $Pr=\nu/\nu_{th}$ is the thermal  Prandtl number and $K$
stands for a possible volumic heat source. Considering a gravity
$\mathbf{g}=g_{(r,\phi)} g_0 \mathbf{e}_g $, where $\mathbf{e}_g$ is
a unit vector, $(r,\phi)$ the cylindrical coordinates in the
equatorial plane, and $g_0$ the gravity at the radius $R_2$, the
dimensionless buoyancy force to add in the Navier-Stokes equations
using the Boussinesq approximation is $\mathbf{f_B} =\tilde{Ra}\
\theta\ g_{(r,\phi)}\ \mathbf{e}_g$, with the modified Rayleigh
number $\tilde{Ra}=\vartheta\ [\theta_1-\theta_2] g_0 / \Omega^2
{R_2}$. In a planetary context, the temperature contrast
to take into account only corresponds to the non-adiabatic component, which is the deviation from the thermodynamical equilibrium
state.

We also take the possible presence of an uniform
imposed magnetic field $B_0$ along the rotation axis
$\mathbf{e_{x_3}}$ into account, which is used as the magnetic field scale. The magnetohydrodynamic (MHD) equations then have to be solved simultaneously
\begin{eqnarray}
\nabla \cdot \mathbf{B} &=& 0 \label{eq:divB} \\
\frac{\partial \mathbf{B}}{\partial t} +(\mathbf{u} \cdot \nabla)\
\mathbf{B} &=& (\mathbf{B} \cdot \nabla)\ \mathbf{u}+\frac{1}{Rm}\
\nabla^2\mathbf{B} \label{eq:B}
\end{eqnarray}
with the magnetic Reynolds number $Rm=\sigma_e\ \mu\ \Omega
{R_2}^2$. The magnetic Lorentz force acting on the flow is given by
$\mathbf{f_L}=(\Lambda/Rm)\ (\nabla \times \mathbf{B}) \times
\mathbf{B}$, with the Elsasser number $\Lambda=\sigma_e\
B_0^2/(\rho_0\ \Omega)$. All dimensionless parameters are listed in
Table \ref{tableNSD}.

\begin{table}
 \caption{List of relevant dimensionless parameters.}\label{tableNSD}
 \begin{tabular}{@{}lll}
 Aspect ratio of the shell & $\eta$\\
 Ellipticity of the distortion &  $\beta=(a_{1}^2-a_{2}^2)/(a_{1}^2+a_{2}^2)$  \\
 Distortion rotation rate & $\gamma_{(t)}$\\
 Orbital eccentricity & $e$\\
 Physical libration rate & $\omega_{(t)}$\\
 Physical libration amplitude & $\epsilon$ \\
 Volumic heat source & $K$\\
 Ekman number & $E=\nu/(\Omega\ {R_2}^2)$   \\
 Thermal Prandtl number & $Pr=\nu/\nu_{th}$ \\
 Magnetic Prandtl number & $Pm=\sigma_e\ \mu\ \nu$ \\
 Modified Rayleigh number & $\tilde{Ra}=\vartheta\ [\theta_1-\theta_2] g_0 / \Omega^2 {R_2}$    \\
 Magnetic Reynolds number & $Rm=\sigma_e\ \mu\ \Omega {R_2}^2$    \\
 Elsasser number & $\Lambda=\sigma_e\ B_0^2/(\rho_0\ \Omega)$   \\
 \end{tabular}

\end{table}

\subsection{Base fields}\label{sec:keplerian}

In the reference frame where the elliptical deformation is
stationary, the differential rotation of the fluid has an amplitude
$1-\gamma_{(t)}$. Besides this, the ellipticity induces an elongational
flow $-(1-\gamma_{(t)})\ \beta\ (x_2\ \mathbf{e_{x_1}}+ x_1\
\mathbf{e_{x_2}}) $, leading to the general elliptical base flow
\begin{eqnarray}
\mathbf{U}=(1-\gamma_{(t)})\ [-(1+\beta) x_2
\mathbf{e_{x_1}}+(1-\beta) x_1 \mathbf{e_{x_2}}] \label{baseflow}
\end{eqnarray}
This flow represents the laminar response of the fluid to the tidal
distortion as an exact, non-linear solution of Navier-Stokes
equations for any finite viscosity, provided that $(\nabla \times
\mathbf{f}) \cdot \mathbf{e_3}=2\ \textrm{d}_t \gamma(t) $. This
means that a body volumic force noted $\mathbf{f}$ is necessary to have spin
period fluctuations (i.e. free or forced physical librations), as is obvious in a planetary context. Equation (\ref{baseflow}) leads to elliptical
streamlines of instantaneous ellipticity $\beta$. We note that $\beta$
is not the mathematical eccentricity of the streamlines, given by
$\sqrt{2\ \beta/(1+\beta)}$. Also the velocity magnitude changes
along a streamline, where the isovalues of the velocity are elliptical
but with an ellipticity $2\ \beta$.

We further assume that a stationary temperature profile $\Theta(r,
\phi)$ is imposed, which is, at order 1 in $\beta$, the solution of the energy conservation equation
(\ref{eq:theta}) with the base field $\mathbf{U}$. We suppose that the modified Rayleigh number $\tilde{Ra}$
is such that $\tilde{Ra}=O(\beta)$. We also consider the presence of an imposed
uniform magnetic field along the rotation axis, produced
for instance by a companion body. We assume that the Lorentz force does not modify the base flow but
only plays a role on the elliptical instability. This implies that
this force is $O(\beta)$. In this context, we see below
that, regarding the elliptical instability, equations for fluid
motions on the order of $0$ in $\beta$ are similar to those in the purely
hydrodynamical case, where the elliptical instability is described as
a resonance between two inertial waves. The magnetic and thermal
fields only induce a correction in the fluid equations on the order of $1$ in
$\beta$, hence a correction of the growth rate of the instability,
because of the stabilizing effect of the Lorentz and buoyancy
forces.

\subsection{The WKB method: stability along a
streamline}\label{validation} \label{sec:stability_analysis}

Our local approach is based on the short--wavelength
Lagrangian theory developed in \cite{bayly1986three}, \cite{craik1986evolution}, then
generalized in \cite{friedlander1991instability} and \cite{lifschitz1991local}. This
method has been successfully applied to the elliptical instability
by \cite{le2000three}, then extended to take the
energy equation and the buoyancy force into account \cite[][]{le2006thermo}, or the induction equation
and the Lorentz force \cite{herreman2009effects}. To summarize, the WKB (Wentzel-Kramers-Brillouin) method consists of looking for a
perturbed solution of the equations of motion under the form of
localized plane waves along the streamlines of the base flow. We
thus look for a solution of the linearized non-dimensional system of
equations (\ref{eq:div}, \ref{Navier}, \ref{eq:theta}, \ref{eq:divB}, \ref{eq:B})
in the form
\begin{eqnarray}
\mathbf{u}_{(\mathbf{x},t)} = \mathbf{U} + \mathbf{u'}(t)\ e^{\I \mathbf{k}(t)\cdot\mathbf{x}},\\
\mathbf{p}_{(\mathbf{x},t)} = \mathbf{P} + \mathbf{p'}(t)\ e^{\I \mathbf{k}(t)\cdot\mathbf{x}},\\
\theta_{(\mathbf{x},t)} = \Theta + \theta'(t)\ e^{\I \mathbf{k}(t)\cdot\mathbf{x}},\\
\mathbf{B}_{(\mathbf{x},t)}= \mathbf{\,B_0} + \mathbf{b}(t)\ e^{\I
\mathbf{k}(t)\cdot\mathbf{x}},
\end{eqnarray}
along the streamlines of the base flow described by
\begin{equation}
\frac{\D \mathbf{x}}{\D t}=\mathbf{U}, \label{eq:streamlines}
\end{equation}
where $\mathbf{k}_{(t)}$ is the time-dependent wave vector,
$\mathbf{x}$ the position vector, and where $\mathbf{U}$ (with its
corresponding pressure field $\mathbf{P}$), $\Theta$ and
$\mathbf{\,B_0} = (0,0,1)$ are the dimensionless base fields defined
in section \ref{sec:keplerian}. Dropping the primes for
simplicity, the linearized system of equations writes as
 \begin{eqnarray} \label{linsyst}
\mathbf{k} \cdot \mathbf{u} &=& 0\\
 \D_t \mathbf{u} &+& \I\ \mathbf{u}\ ( \D_t  \mathbf{k} \cdot \mathbf{x}) +\I\  ( \mathbf{U} \cdot \mathbf{k})\ \mathbf{u}+ (\mathbf{u} \cdot \nabla)\ \mathbf{U} +2\ \gamma_{(t)}\ \mathbf{e_{x_3}} \times  \mathbf{u} \nonumber\\ %[3mm]
 &=& - \I\ p\ \mathbf{k}  -k^2 E\ \mathbf{u} +\frac{\Lambda}{Rm}\ (\I\  \mathbf{k} \times \mathbf{b}) \times \mathbf{B_0} + \tilde{Ra}\ \theta\ g\
\mathbf{e}_g\\ \label{eq:NS}
 \D_t \theta  &+& \I\ \theta\  ( \D_t  \mathbf{k} \cdot \mathbf{x}) +\I\  ( \mathbf{U} \cdot \mathbf{k})\ \theta + (\mathbf{u} \cdot \nabla)\ \mathbf{\Theta} = -k^2\frac{E}{Pr}\theta \label{eq:WKB_temp}\\
 \mathbf{k} \cdot \mathbf{b} &=& 0 \\
 \D_t \mathbf{b} &+& \I\ \mathbf{b}\ ( \D_t  \mathbf{k} \cdot \mathbf{x}) +\I\  ( \mathbf{U} \cdot \mathbf{k})\ \mathbf{b} \nonumber\\ %[3mm]
 &=& (\mathbf{b} \cdot \nabla)\ \mathbf{U}+\I\ (\mathbf{B_0} \cdot \mathbf{k})\ \mathbf{u} -\frac{k^2}{Rm}\ \mathbf{b}. \label{eq:WKB_magn}
 \end{eqnarray}

Those equations can be decoupled in space and time to give an
equation for the wave vector only:
\begin{eqnarray}
\D_t \mathbf{k} \cdot \mathbf{x}+\mathbf{U} \cdot \mathbf{k}=0.
\label{eq:wave}
\end{eqnarray}
The solution of the remaining equations for $\mathbf{u}, \theta,
\mathbf{b}$ is then sought under the form of a Taylor expansion in
$\beta$ of all variables, as illustrated in appendices \ref{WKB1} and
\ref{WKB2}.

This approach is used in the following sections to calculate
the growth rate of the instability in the two generic cases: the
TDEI, which appears in the case of non-synchronized bodies, and the
LDEI, which appears in the case of synchronized bodies.

\subsection{Non-synchronized bodies: inviscid growth rate of the TDEI}\label{sec:background}

In this section, we consider the effects of dynamic tides of
amplitude $\beta$ on the liquid core of a Mercury-like planet
orbiting close to its star with (i) constant but different orbital
and sidereal rotation periods, (ii) an imposed thermal
stratification \cite[see e.g.][]{manglik2010dynamo}, and (iii) an
externally imposed magnetic field (e.g. the Sun magnetic field).
The same analysis applies to the stratified zone of a star
(the so-called radiative zone) tidally deformed by a companion body,
taking the magnetic field generated by dynamo in its
convective zone into account. The present configuration corresponds to the
standard case of the elliptical instability as already known, but
completed by the complexities present in real astrophysical cases.
These additive effects have already been studied
separately, even if they are simultaneously present in real systems.
The effect of the angular velocity of the tidal bulge has been
studied in \cite{miyazaki1992three}, \cite{le2000three},
\cite{le2007coriolis} and \cite{le2010tidal} and the presence of a
thermal field has been studied in \cite{le2006thermo}, who study the linear competition between the growths of the TDEI and the convection, as well as in \cite{cebron2010tidal} and \cite{lavorel2010experimental}, who study the growth of the TDEI over established convective flows. The presence of an inner solid core has been studied in \cite{lacaze2005elliptical} and of an external magnetic field in \cite{kerswell1994tidal}, \cite{kerswell2002elliptical}, \cite{lacaze2006magnetic} and \cite{herreman2009effects}. We extend these works by including all of these features in a single formula.

In the non-synchronized case, which is considered in this section, the base flow (\ref{baseflow}) reduces to
\begin{eqnarray}
\mathbf{U}=(1-\Omega_{orb}/\Omega_{spin})\ [-(1+\beta) x_2
\mathbf{e_{x_1}}+(1-\beta) x_1 \mathbf{e_{x_2}}].
\end{eqnarray}
The WKB analysis is then tractable (see appendix \ref{WKB1}), taking thermal and magnetic effects into account in the limit where
buoyancy and Lorentz forces are on the order of $\beta$. The
instability does not exist in the range $\Omega_{spin}/\Omega_{orb}
\in [-1;1/3]$, which is called the forbidden band. It corresponds to
the absence of resonance between the elliptical forcing and the
inertial waves of the rotating flow \cite[see][for a complete
discussion]{le2007coriolis}. In the present limit, the presence of
the thermal and magnetic fields does not affect the forbidden band.
Neglecting the thermal diffusion, the inviscid growth rate of the
TDEI with the presence of thermal and magnetic fields is
\begin{eqnarray}
\displaystyle \sigma_{inv} =  \frac{\sqrt{(2\Omega^G+3)^4\ \beta^2-4\ \left(\tilde{Ra}\ r\ \partial_r \Theta \right)^2 }}{16\ |1+\Omega^G|^3} -\frac{\Lambda}{4\ |1+\Omega^G|^3}, \label{eq:general}
\end{eqnarray}
with $\Omega^G=\Omega_{orb}/(\Omega_{spin}-\Omega_{orb})$, $r$ the radius and
$\partial_r \theta$ the
dimensionless temperature base-field radial gradient on the
considered streamline. This expression allows to recover the
different cases already obtained in the literature. For instance,
the purely hydrodynamic growth rate given by
\cite{miyazaki1992three}, \cite{le2000three}, and \cite{le2010tidal}
is recovered for $(\tilde{Ra}=0,\ \Lambda=0)$. For a fixed
elliptical deformation, we recover the classical inviscid value
$\sigma_{inv}/\beta=9/16$. Finally, in the absence of a thermal
field and with a stationary bulge $(\Omega_{orb}=0,\ \tilde{Ra}=0)$,
the magnetic case given in \cite{herreman2009effects} is also
recovered. Formula (\ref{eq:general}) is fully generic and clearly
illustrates the stabilizing influence of Joule dissipation and of a
local stratification in the range of validity of this stability analysis (see section \ref{sec:validity_approach} and appendix \ref{GI}).

\subsection{Synchronized bodies: inviscid growth rate of the LDEI} \label{sec:orblib}

In this section, we consider the liquid ellipsoidal core of a
synchronized moon like Io or of an extrasolar telluric planet
orbiting close to its massive attractor with (i) an orbital period
equal to the sidereal rotation period, but with small instantaneous
fluctuations of the differential rotation between the elliptical
deformation and the fluid, whatever their origin (optical or
physical, forced or free libration, longitudinal flows induced by
tides); (ii) an imposed magnetic field (e.g. Jupiter's magnetic
field for Io); and (iii) a local thermal gradient. The dimensionless
instantaneous differential rotation between the fluid and the
elliptic deformation oscillates with a libration amplitude
$\epsilon$ (equal to $2e$ for optical librations) and a libration
frequency $\omega_o$ (equal to $1$ for forced librations).
Considering the particular case of a fluctuation due to the orbital
ellipticity (i.e. $\omega_o=1$), \cite{kerswell1998tidal} and
\cite{herreman2009effects} have shown that these oscillations can
lead to LDEI. We extend these previous studies to the more general
case of small fluctuations of arbitrary periods, taking buoyancy into account. Neglecting the thermal diffusion, the inviscid growth
rate of the LDEI is then (see appendix \ref{WKB2})
\begin{eqnarray}
\sigma_{inv} = \frac{16+\omega_o^2}{64}\ \sqrt{(\epsilon\ \beta)^2-\frac{4}{\omega_o^2}\ \left(\tilde{Ra}\ r\ \partial_r \Theta \right)^2} -\frac{\omega_o^2}{16}\ \Lambda \label{eq:rate}
\end{eqnarray}
at first order in $\epsilon \beta$, taking the effects
of thermal and magnetic fields into account in the limit where buoyancy and
Lorentz forces are $O(\beta)$. The forbidden band is given
by $|\omega_o| > 4 $. As before, the generic formula (\ref{eq:rate}) clearly illustrates the
stabilizing influence of Joule dissipation and of a local
stratification.

The case $\omega_o \rightarrow 0$ corresponds to the limit toward
the fully synchronized state. In the case of the TDEI, this limit
case is obtained with $\Omega^G \rightarrow \infty$, which gives the
inviscid growth rate $\sigma_{inv}=\beta/(4 \Omega^G)$ for large
wavenumbers. Both expressions for the growth rate are thus
consistent in the limit of synchronized state:
$\Omega_{orb}/\Omega_{spin}=1-\epsilon$ i.e. $ |1+\Omega^G| \sim
|\Omega^G|=1/\epsilon $. The expression given in
\cite{herreman2009effects} is also exactly recovered when $\omega_o
= 1$ and $\Lambda=0$. There is a slight error on the
numerator of the magnetic damping term in \cite{herreman2009effects}: in their considered case, $\omega_o = 1$, the magnetic damping is erroneously
$  -\Lambda / 16$, instead of  $ -3 \Lambda/16$.

\subsection{Viscous dissipation} \label{damping_term}

The previous sections present the calculation of the growth rate of
the elliptic instability in an inviscid fluid with Joule dissipation
and buoyancy stabilization. Calculation of the threshold of the
instability requires correctly estimating all dissipative
terms. In the case of no-slip boundaries, dissipation occurs mainly
in the viscous boundary layers of thickness $E^{1/2}$. This implies
a damping term that should alter the growth rate:
\begin{eqnarray}
\sigma=\sigma_{inv}-\alpha\ E^{1/2}\ f(\eta), \label{visccorrect}
\end{eqnarray}
where $\alpha$ is a constant between $1$ and $10$, equal to $\alpha=2.62$ and $f(\eta)=(1+\eta^4)/(1-\eta^5)$  for
the spinover mode of the TDEI \cite[see e.g.][]{kudlick1966transient,hollerbach1995oscillatory,lacaze2005elliptical}.

In addition to decreasing the growth rate, viscous dissipation is
also primordial for quantifying the orbital evolution and
rotational history of a binary system during its synchronization. A
model has been proposed in \cite{le2010tidal} for $\Omega_{orb}=0$,
which allows the authors to estimate the viscous power dissipated by
TDEI. Our purpose here is to generalize this model to all cases
studied above. Far from threshold, the model proposed by
\cite{le2010tidal} considers that the TDEI simply corresponds to a
differential rotation between the boundary and the bulk. According
to this model, the power dissipated by the system is
\begin{eqnarray}\label{scalingpower}
P= -2\ M\ R_2^2\ {\bigtriangleup \Omega}^2\ \Omega E^{1/2},
\end{eqnarray}
assuming that in the small Ekman numbers limit reached in
astrophysical cases, the amplitude of the instability is
commensurate with the differential rotation $\bigtriangleup \Omega$
\cite[][]{cebron2010systematic}.

The tidal quality factor $Q$ is widely used in systems evolution
calculations. By analogy with the theory of harmonic
oscillators, $Q$ is defined by \cite[e.g.][for a recent discussion on $Q$]{greenberg2009frequency} the ratio between the maximum
potential gravitational energy stored in the tidal distortion over
the energy dissipated in one work cycle. The dissipated power associated to the flow driven by the elliptical instability does not have the periodicity of the forcing, so a quality factor cannot be rigorously defined in the same way. However, we can define a closely related ratio $Q^*$, comparing the power
typically dissipated in the fluid layer over one revolution with the
potential energy stored in the bulge, which is on the order of $E_0 \sim 4 \pi \rho_0\ g_0\ s^2\ R_2^2$,
with $s$ the dimensional height of the tides
\cite[e.g.][]{benest1990modern}. Since $\beta \sim s/R_2$, we obtain
from equation (\ref{scalingpower})
\begin{eqnarray}
Q^* \sim \frac{g_0\ \beta^2}{R\bigtriangleup \Omega ^{2}\ E^{1/2}}.
\end{eqnarray}
The ratio $Q^*$ gives a dimensionless measure of the strength of the
dissipation in the fluid.

\subsection{Validity of the approach} \label{sec:validity_approach}

The previous analysis is valid when the elliptic instability
comes from a resonance of pure hydrodynamic inertial waves.
Therefore, any previously derived expressions are limited to the
case where buoyancy and Lorentz forces are $O(\beta)$. According to (\ref{linsyst}), this
means that
\begin{eqnarray}
\frac{\Lambda}{Rm}\ k \sim \beta \quad \mbox{and} \quad \tilde{Ra}
\sim \beta,
\end{eqnarray}
where $k$ is the dimensionless norm of the wavevector of the excited
mode. For a typical planetary core, these conditions can be
rewritten
\begin{eqnarray}
B_0 \sim 0.1 \frac{\sqrt{\beta/k}}{R_2 E}\quad \mbox{in}\ \mu\mbox{T}\quad \mbox{and} \quad \frac{F_{non-adia}}{F_{adia}} \sim
10^{-3}\ \frac{\beta}{E^2 R_2^4 g_0^2}, \label{conditions}
\end{eqnarray}
where $F_{non-adia}$ and $F_{adia}$ are the
non-adiabatic and adiabatic components of the core heat flux, respectively. The
condition on the magnetic field is easily verified for planetary
cores over a wide range of wave vector $k$. The condition on the
non-adiabatic heat flux is more problematic to quantify: in most
planets, the adiabatic profile is supposed to be sufficient to
transport core heat flux, and the non-adiabatic component is
estimated to be very small, but it is not known precisely. One
should notice that the condition (\ref{conditions}) is very
restrictive, and special attention should be paid in each given
configuration. For instance, in the case of Europa, Eq. (\ref{conditions}) implies $F_{non-adia}/F_{adia} \sim 0.1 \% $,
which seems reasonable; but in the case of Io, it implies $ F_{non-adia}/F_{adia} \sim 1 \% $, which is only
marginally verified since the estimated non-adiabatic heat flux is
about one fifth of the adiabatic component
\cite[e.g.][]{kerswell1998tidal}.

Now supposing that the buoyancy or the Lorentz force is on the order of $0$
in $\beta$, we can wonder if the elliptic instability still exists.
In this case, inertial waves are replaced by gravito-inertial or
magneto-inertial waves, and the elliptical instability arises as a
resonance between those modified waves. Resonances of
magneto-inertial waves has been studied, for instance, in
\cite{kerswell1994tidal}, \cite{lebovitz2004magnetoelliptic} and \cite{mizerski2009magnetoelliptic}
in the case of an imposed uniform magnetic field along the spin
axis. This is the so-called magneto-elliptic instability. Resonance
of gravito-inertial waves has been studied in
\cite{le2006thermo} and \cite{guimbard2010elliptic}, who concluded that a
stratified field can either be stabilizing or destabilizing
depending on the shape of gravitational iso-potentials and isotherms: the so-called gravito-elliptic instability. This point is
further clarified in appendix \ref{GI}, which shows the high
sensitivity of the elliptical instability to the specificities of
the thermal and gravity fields. Planets with buoyancy or Lorentz
force on the order of $0$ in $\beta$ should be the subject of specific
studies, which is beyond the scope of the generic results presented
here.

\section{Application to solar/extrasolar systems} \label{sec:applications}

Using previous results, we are now in a position to calculate the
threshold of the elliptical instability for telluric bodies of different
systems. We consider that a body is stable or unstable when the mean value of the growth rate
over an orbit is either positive or negative.

\subsection{Non-synchronized system: tide-driven elliptical instability (TDEI)}

\begin{table}
\caption{Physical and orbital characteristics used for the stability
calculations and results.}

\label{table_solar}
 \begin{tabular}{@{}lcccccc}
     & Mercury & Venus
        & Earth & Early Earth$^b$   \\
$M$ ($\times 10^{-24}$ kg) & $0.330$ & $4.87$ & $5.98$ & $5.98$   \\
  $R$ (km) & $2440$ & $6051$ & $6378$ & $6378$ \\
 $T_{spin}\  (d)$ & $58.6$ & $-243$ & $0.997$ & $0.418$  \\
 $T_{orb}\  (d)$ & $87.97$ & $224.7$ & $27.32$ & $9.67^b$ \\
 Tidal amplitude (m) & $0.925^a$ & $1.8$ & $0.6$ & $3.4$ \\
 $R_2/R$ & $0.8^a$ & $0.17$ & $0.55$ & $0.55$ \\
 $\eta$ & $0$ & $0$ & $0.35$ & $0$ \\
 $E$ ($\times 10^{14}$) & $21$ & $316$
        & $0.11$ & $0.047$\\
$\beta$ (measured) ($\times 10^{7}$) & $7.6$ & $?$ & $1.9$ & $11$\\
$\beta$ (hydrostatic) ($\times 10^{7}$) & $6.8$ & $1.1$
        & $0.8$ & $6.7$\\

${B_{suf}}^c$ (nT) & $250^d$ & $30$ & $3 \cdot 10^{4}$ & $0$ \\
${B_0}^e$ (nT) & $488$ & $6100$ & $1.8 \cdot 10^{5}$ & $0$ \\
$\Lambda$  & $6.4 \cdot  10^{-6}$ & $0.004$ & $0.015$ & $0$\\
$\sigma\ (years^{-1})$ & $-1.5 \cdot 10^{-5}$ & $-5.43$ & $-7.7$ & $0.003$\\
 \smallskip
 \end{tabular}\\

Following \cite{herreman2009effects}, we take as
typical values $\sigma_e=4 \cdot 10^{5}\ \textrm{S.m}^{-1}$, $\rho_0=12\
000\ \textrm{kg.m}^{-3}$ and $\nu=10^{-6}\ \textrm{m}^2. \textrm{s}^{-1}$, consistent with a
Fe/Fe-S composition.\\

$^a$ \cite{van2007mercury}\\
$^b$ Considering an Early Moon two times closer than today.\\
$^c$ Equatorial surface field\\
$^d$ \cite{anderson2010magnetic}\\
$^e$ Considering a variation in $r^{-3}$ from the core to the planetary surface ($r$ being the spherical radius).\\
\end{table}

We consider first the TDEI in liquid cores of telluric bodies in
the solar system. A rough criteria given by the equations
(\ref{eq:general}) and (\ref{visccorrect}) leads to a threshold
$\beta/\sqrt{E} \sim O(1)$, as already mentioned. In the solar
system, this leads to focus only on Mercury, Venus, and on the
Earth-Moon system during its evolution. Most tidal evolution models
predict that the Moon rapidly retreats to $25-35$ Earth radii in less
than about $100\ \textrm{Ma}$
\cite[][]{webb1982tides,ross1989evolution,williams2000geological,williams2004paradox},
and we thus consider two limit cases: the actual Earth-Moon system
and an early Earth with an early Moon at $30$ Earth radii, i.e. two
times closer than today. The tabulated values found in the
literature for these planets are given in Table \ref{table_solar}. The Ekman number $E$ is calculated with a molecular kinematic viscosity $\nu=10^{-6}\ \textrm{m}^2. \textrm{s}^{-1}$,
consistent with Fe/Fe-S composition of a liquid outer core.

\begin{figure}
  \begin{center}
    \epsfysize=6.0cm
    \leavevmode
    \epsfbox{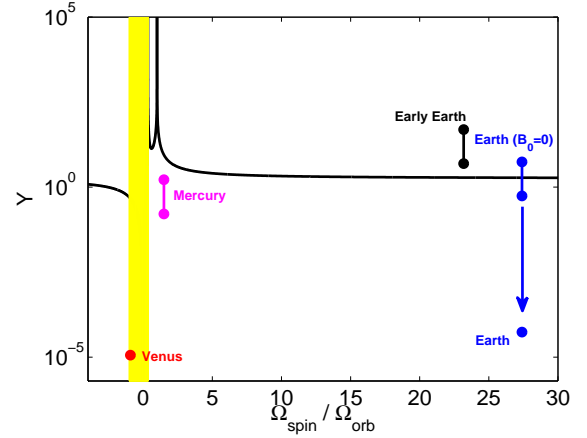}
    \caption{TDEI stability diagram for celestial bodies of the solar system.
    Considering a surfacic viscous damping term of the growth rate $\sigma_s=-\alpha\ f(\eta)\ E^{1/2}$
    (see section \ref{damping_term}), the zone above the black line, defined by (\ref{eq:figTDEI}),
    is the unstable zone, whereas $Y$, defined by (\ref{eq:TEDI_Y}), is calculated with $\alpha=1$ and $\alpha=10$ for each planet (for the actual Earth and Venus, the difference is very small). The yellow zone is the so-called 'forbidden zone', given by $\Omega_{spin}/\Omega_{orb} \in [-1;1/3]$.}
    \label{cebronfig4}
  \end{center}
\end{figure}

We first neglect thermal effects. To represent all bodies
of Table \ref{table_solar} on the same stability diagram, we define
the quantity
\begin{eqnarray}
Y= \beta \left ( \alpha f(\eta) \sqrt{E}+\frac{\Lambda}{4 |1+\Omega^G|^3}\
\right)^{-1}. \label{eq:TEDI_Y}
\end{eqnarray}
The quantity $Y$ includes the specific dependence of the growth rate on the spin/orbit angular
velocity ratio, aspect ratio $\eta$ of the inner core and the magnetic field. The
stability criterium
\begin{eqnarray}
\sigma= \frac{(2\Omega^G+3)^2}{16\ |1+\Omega^G|^3}\ \beta - \alpha\ f_{(\eta)} \sqrt{E} -\frac{\Lambda}{4\ |1+\Omega^G|^3}  \geq 0  \label{stab_crit}
\end{eqnarray}
derived from (\ref{eq:general}) and (\ref{visccorrect}) is then equivalent to
\begin{eqnarray}
Y \geq \frac{16\ |1+\Omega^G|^3 }{(3+2 \Omega^G)^2}. \label{eq:figTDEI}
\end{eqnarray}
Figure \ref{cebronfig4} represents the stability results for the
TDEI in the liquid cores of the non-synchronized planets of the
solar system considered in Table \ref{table_solar}. The case of each
planet is discussed in the following.

An important result, already noticed in \cite{cebron2010systematic},
is that the Early Earth, with a Moon two times closer than today and
in the absence of an external magnetic field, is unstable with a good
level of confidence. The dissipated power due to the instability was
around $5 \cdot 10^{18}\ W$, which corresponds to $Q^* \sim 0.003$. This estimation seems huge in comparison
to the present dissipation by tidal friction \cite[$\sim3.75 \cdot
10^{12}\ W$ according to][]{munk1998abyssal}. However, one
must notice that the estimations given here are based on a model
where the rotation rate is implicitly expected to be constant,
corresponding to a quasi-static approximation of the orbital
evolution of the system. This approximation obviously breaks down
for a high dissipation rate, and the above result should be
interpreted as proof of a rapid orbital evolution of the
Earth-Moon system. The Moon orbit inclination is not taken
into account in the stability analysis considered in this work. A
similar analysis would be difficult because the forced base flow in
such a configuration is not analytically known. However, it has been
demonstrated numerically in \cite{cebron2010systematic} that this inclination only has minor consequences on the process of elliptical instability, which may be easily taken into account in considering
an effective elliptical distortion in the equatorial plane. In the
present case, this orbital inclination would slightly decrease our
growth rates but does not change the orders of magnitude our
conclusions either.

\begin{figure}
  \begin{center}
    \epsfysize=6.0cm
    \leavevmode
    \epsfbox{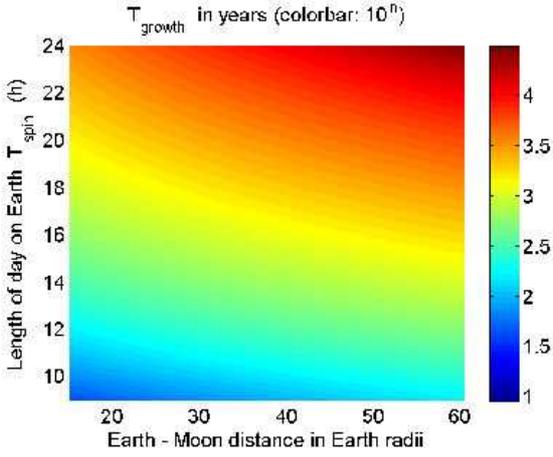}
    \caption{Evolution of the typical growth time $T_{growth}=1/ \sigma$ of the instability in the Early Earth core as a function of the Earth-Moon separation and of
    the length of day on the Early Earth $T_{spin}$ ($\alpha=2.62$). This diagram assumes no external magnetic field ($B_0=0$) and no thermal field ($\tilde{Ra}=0$).}
    \label{fig:early}
  \end{center}
\end{figure}

Figure \ref{fig:early} shows the stability of the Early Earth in more detail, for different values for the length of day and Earth-Moon distances. In the absence of meteoroid impacts, the
angular momentum conservation links these two quantities, which
cannot vary independently. However, at this epoch, violent meteoroid
impacts have probably modified the angular momentum of the early
Earth-Moon system \cite[][]{melosh1975large,wieczorek2009did}, so we keep these two parameters independent, which allows us to cope with uncertainties.

The case of the actual Earth is more subtle because if we consider that the
actual magnetic field is provided by thermo-solutal convective
motions in the core, it has to be considered as an imposed field for
the dynamics of the elliptical instability. In this case, the
destabilizing term in the growth rate (\ref{stab_crit}) is about
$10^{-7}$, whereas the magnetic damping term is around $0.004$. Then
the TDEI cannot grow, regardeless of the Ekman number. On the contrary,
if we consider that the actual magnetic field is provided by the
flow driven by the TDEI, the threshold has to be calculated with
$\Lambda=0$, and the actual Earth is slightly unstable, with a
growth time of around 14,000 years. The same result was suggested by
\cite{aldridge1997elliptical}, but neglecting the influence of magnetic
field and global rotation.

We now consider the influence of thermal effects. Considering
the actual heat flux of Earth, \cite{christensen2006scaling}
estimate the corresponding super-adiabatic temperature contrast to
be about $1\ \textrm{mK}$, leading to a vigorous convection in the
liquid core. Such a vigorous convection does not prevent the
elliptical instability from growing, as shown in
\cite{cebron2010tidal,lavorel2010experimental}. Since this
temperature contrast is uncertain, we can consider as an extreme
case for stabilization of the TDEI, a subadiabatic gradient on the
same order of magnitude. For the actual Earth, such a stratification
leads to $\tilde{Ra} \approx -1.4 \cdot 10^{-6}$, and the dependence
$\tilde{Ra} \propto D^3\ E^2$ gives the value $\tilde{Ra} \approx
-2\cdot 10^{-7}$ for the Early Earth \cite[see][for a discussion of
this possible stable density stratification in the whole early Earth
outer core and its disruption]{surnita2003thermal}. Considering
$\tilde{Ra} \approx -1 \cdot 10^{-6}$ as an upper bound, formula
(\ref{eq:general}) is valid ($\tilde{Ra} \sim O(\beta)$) and shows
that the thermal stratification reduces the growth rate by $2 \%$.
This confirms that the role of the temperature can be neglected in
the limits considered in this work.

For the last two bodies, Venus is in the forbidden band,
which means that no matter what the tidal deformation or the Ekman number
is, the TDEI cannot grow. Mercury is slightly below the threshold
of the instability and is thus probably stable today. Mars
is clearly stable nowadays ($\beta \ll \sqrt{E}$), but
\cite{arkani2008tidal} and \cite{arkani2009did} suggest that past
gravitational interactions with asteroids could have excited a TDEI
in the martian core during their fall towards the planet.

\subsection{Synchronized body: libration-driven elliptical instability (LDEI)}

\subsubsection{Galilean moons and Titan}

\begin{table}
\caption{Physical and orbital characteristics used for the four
Galilean moons and Titan.}

\label{table_Io}
 \begin{tabular}{@{}lcccccc}
      & Io & Europa
        & Ganymede & Callisto & Titan \\
 $M$ ($\times 10^{-22}$ kg) & $8.93$ & $4.8$ & $14.8$ & $10.8$ & $13.45$  \\
 $R$ (km) & $1822$ & $1561$ & $2631$ & $2410$ & $2576$ \\
 $T_{orb}\  (d)$ & $1.77$ & $3.55$ & $7.16$ & $16.69$ & $15.95$ \\
 $e$ ($\times 10^{3}$) & $4.1$ & $9.4$ & $1.3$ & $7.4$ & $28.8$ \\
 $\epsilon$ ($\times 10^{4}$) & $ 1.3^a$ & $2^a$ & $ 0.056^a$ & $ 0.042^a$ & $1.3^a$ \\
$B_0$ (nT)  & $1850$ & $410^b$ & $120$ & $10^b$ & $0$\\
 \end{tabular}

 \medskip
$^a$ Physical libration amplitude from \cite{noir2009experimental}\\
$^b$ Order of magnitude from \cite{zimmer2000subsurface} of the magnetic field component along the rotation axis of the moon \cite[see also][]{kabin1999europa}. \\
\end{table}

The presence of LDEI in Io has been first suggested by
\cite{kerswell1998tidal}. In \cite{herreman2009effects}, this suggestion is reexamined and the magnetic field induced by this possible instability quantified. In the following, this calculation
is re-evaluated and extended to the four Galilean moons (Io, Europa,
Ganymede, and Callisto), considering the presence of the external
magnetic field of Jupiter. Titan is also considered. All necessary
data are given in Tables \ref{table_Io} and \ref{table_Io2}.

\begin{table}
\caption{Stability results in the extreme case of optical librations
($\epsilon=2e$ and $\beta$ from this Table and Table
\ref{table_Io}).}

\label{table_Io2}
 \begin{tabular}{@{}lcccccc}
      & Io & Europa
        & Ganymede & Callisto & Titan \\
      &  &
        &  &  &  \\
      & core & core
        & core & core & core \\
 $R_2/R$ & $0.52^{a,b}$ & $0.38^b$ & $ 0.27^c$ & $-$ & $-$ \\
 $\eta$ & $0$ & $0$ & $0$ & $-$ & $-$ \\
 $E$ ($\times 10^{14}$) & $2.7$ & $14$
        & $20$ & $-$ & $-$\\
$\beta$ ($\times 10^{4}$) & $60^a$ & $9.7^d$ & $3.7^e$ &
$-$ & $-$\\
$\Lambda$ ($\times 10^{7}$) & $42$ & $4.1$ & $0.7$ & $-$ & $-$\\
$\sigma\ (yr^{-1})$ & $0.016$ & $0.0025$ & $-3 \cdot 10^{-4}$ & $-$ & $-$\\
      &  &
        &  &  &  \\
      & ocean & ocean
        & ocean & ocean & ocean \\
 Crust (km) & $-$ & $10^g$ & $ 100^h$ & $150^i$ & $70^j$ \\
 Depth (km) & $-$ & $100^g$ & $ 150^h$ & $150^i$ & $200^j$ \\
 $R_2/R$ & $-$ & $0.99$ & $ 0.96$ & $0.94$ & $0.97$ \\
 $\eta$ & $-$ & $0.94$ & $0.94$ & $0.93$ & $0.92$ \\
 $E$ ($\times 10^{14}$) & $-$ & $2.0$
        & $1.5$ & $4.5$ & $3.5$\\
$\beta$ ($\times 10^{4}$) & $-$ & $9.7^d$ & $3.7^e$ &
$0.72^e$ & $1.2^f$\\
$\Lambda$ ($\times 10^{13}$) & $-$ & $21$ & $3.5$ & $0.9$ & $0$\\
$\sigma\ (yr^{-1})$ & $-$ & $0.0016$ & $-6 \cdot 10^{-4}$ & $-4 \cdot  10^{-4}$ & $-10^{-4}$\\
 \end{tabular}

 \medskip

For the liquid cores, we take $\sigma_e=4 \cdot 10^{5}\ \textrm{S.m}^{-1}$, $\rho_0=8\ 000\
\textrm{kg.m}^{-3}$ and $\nu=10^{-6}\ \textrm{m}^2.
\textrm{s}^{-1}$ as typical values, consistent with a Fe/Fe-S composition. For the subsurface oceans, we take $\sigma_e=0.25\ \textrm{S.m}^{-1}$ \cite[][]{hand2007empirical},
$\rho_0=1 000\ \textrm{kg.m}^{-3}$, and $\nu=10^{-6}\
\textrm{m}^2.\textrm{s}^{-1}$ as typical values. \\

$^a$ \cite{kerswell1998tidal}, considering the static tidal bulge\\
$^b$ \cite{hussmann2004thermal}\\
$^c$ \cite{bland2008production}\\
$^d$ with $k_2 \approx 0.3$ \cite[][]{wahr2006tides,baland2010librations} \\
$^e$ Eq. (\ref{eq:newton2}) with $k_2 \approx 0.3$ \\
$^f$ $k_2 \approx 1$ \cite[][]{goldreich2010elastic}\\
$^g$ \cite{wahr2006tides}\\
$^h$ \cite{bland2009orbital}\\
$^i$ \cite{kuskov2005internal}\\
$^j$ \cite{sohl2003interior}
\end{table}

\begin{figure}
  \begin{center}
    \epsfysize=6.0cm
    \leavevmode
    \epsfbox{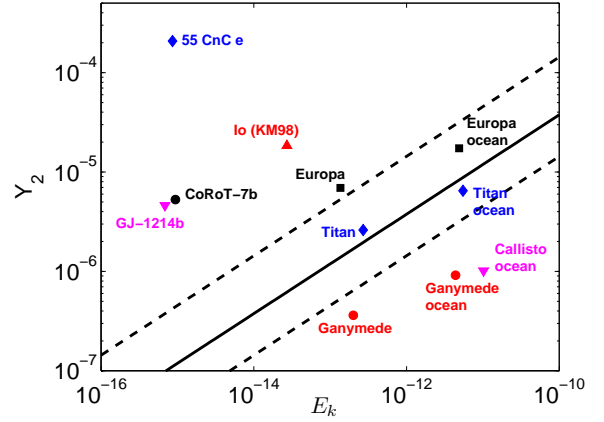}
    \caption{LDEI stability diagram for synchronized celestial bodies in the optimal case for the instability, i.e. an optical libration ($\epsilon=2e$)
    for a quasi-equilibrium tide. All values are given in Table \ref{table_Io}, and the horizontal axis represents the Ekman number $E_k$ based on the thickness of the fluid layer.
    The label KM98 for Io calls that the point is placed with the values used by \cite{kerswell1998tidal}.
    The zone below the black line, defined by the viscous surfacic damping coefficient $\alpha=2.62$, is the stable zone, whereas the black dashed lines represent the extremum values $\alpha=1$ and $\alpha=10$.}
    \label{cebronfig11}
  \end{center}
\end{figure}

\begin{figure}
  \begin{center}
  \begin{tabular}{ccc}
      \setlength{\epsfysize}{5.0cm}
      \epsfbox{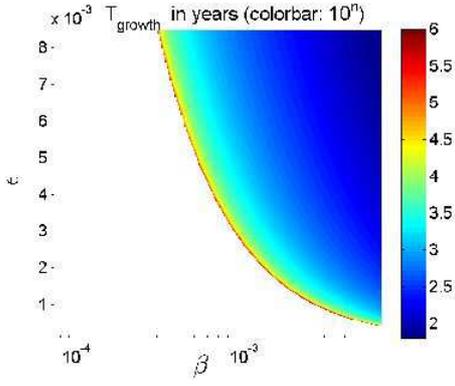} \\
      \setlength{\epsfysize}{5.0cm}
    \end{tabular}
    \caption{Evolution of the typical growth time $T_{growth}=1/ \sigma$ of the instability in Io with the tidal bulge ellipticity $\beta$ and the libration amplitude $\epsilon$ ($\alpha=2.62$). The upper right corner corresponds to the optimal case considered in \cite{kerswell1998tidal} and \cite{herreman2009effects},  the upper left corner corresponds to optical librations with a small tidal amplitude corresponding to diurnal tides ($\beta=7 \cdot 10^{-5},\ \epsilon=2e=0.0082)$ and the lower right corner corresponds to the region of physical librations of the static bulge ($\beta=0.006,\ \epsilon=1.3 \cdot 10^{-4}$). The white zone corresponds to the stable zone where the instability cannot grow because of dissipative effects (at the boundary with the colored zone, the growth time is infinite). The colorbar
    range is chosen so that color variations are visible.}
    \label{fig:Io}
  \end{center}
\end{figure}

\begin{figure}                    % Chaque figure doit avoir pour nom nomfig1.eps,
  \begin{center}
    \begin{tabular}{ccc}
      \setlength{\epsfysize}{5.0cm}
      \subfigure[]{\epsfbox{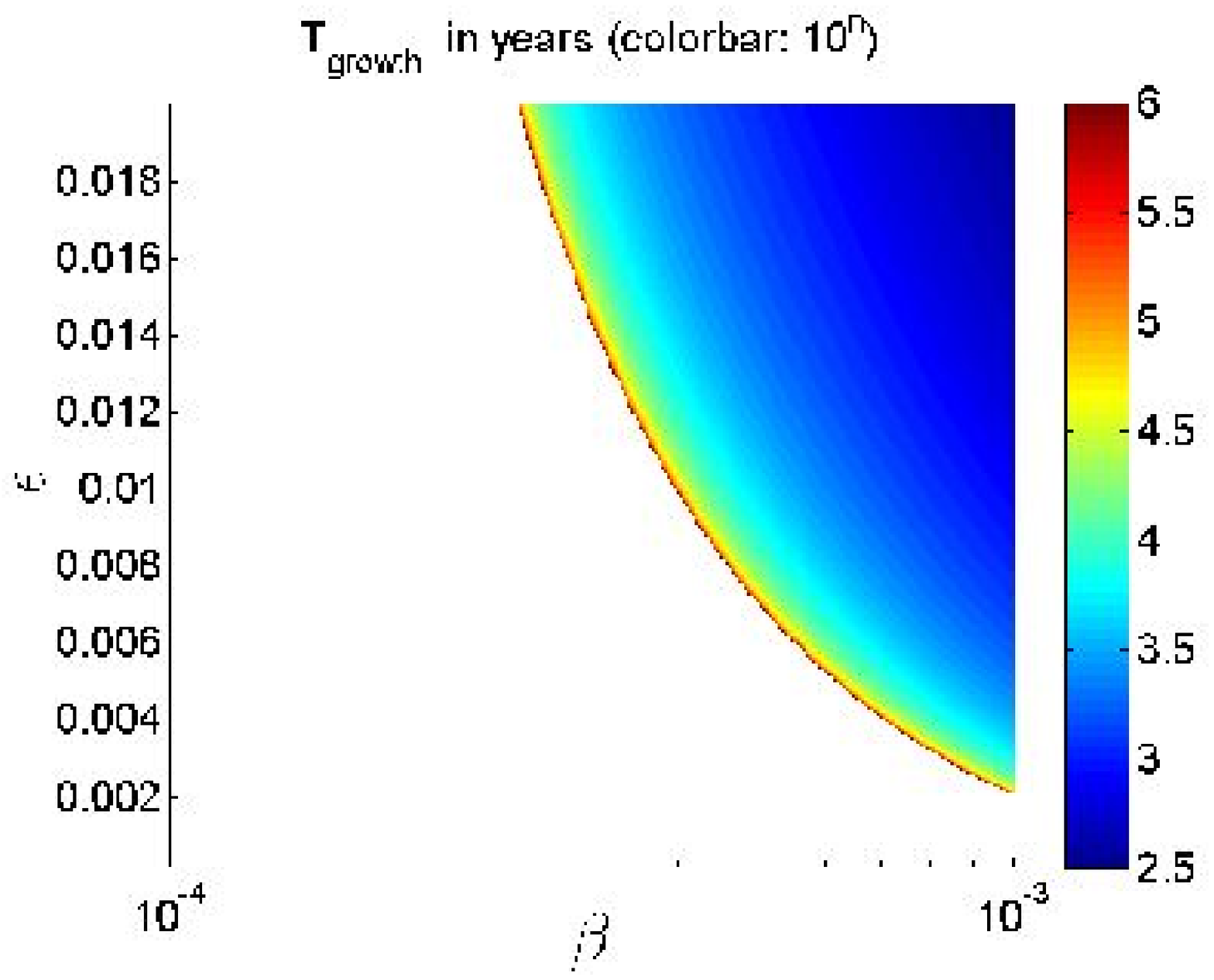}} \\
      \setlength{\epsfysize}{5.0cm}
      \subfigure[]{\epsfbox{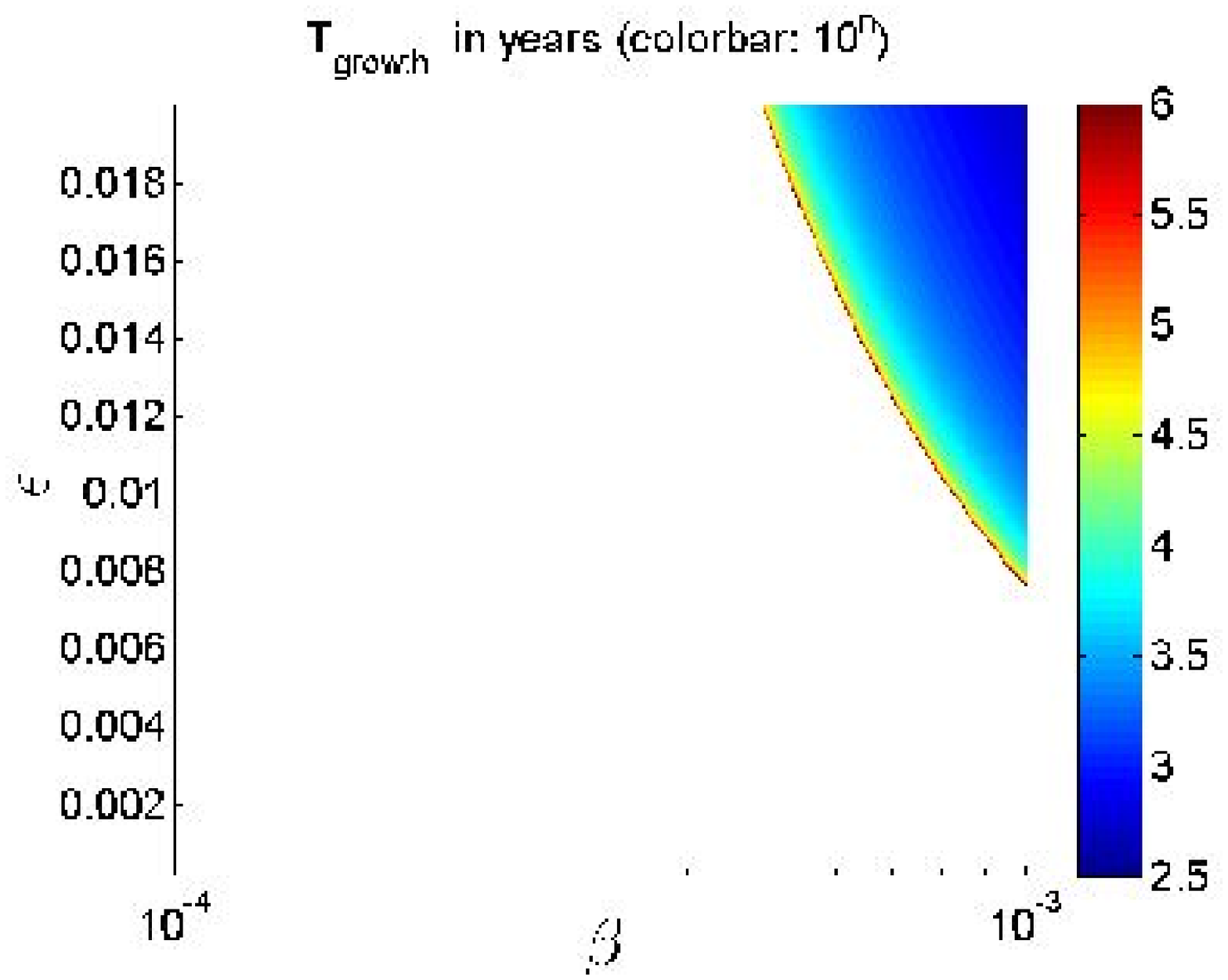}}
    \end{tabular}
    \caption{Same as figure \ref{fig:Io} but for Europa, considering (a) its possible liquid core and (b) a $100$ km depth subsurface ocean. Both the core and the subsurface ocean of Europa are expected to behave in the deformable limit, locating their states in the upper right corner of the diagram, so both are found to be unstable.}
    \label{fig:europa}             %
  \end{center}
\end{figure}

As described in section \ref{sec:orblib}, we consider an
instantaneous differential rotation $\epsilon \cos (\omega_o t)$ for
these synchronized bodies. Focusing on the forced librations due to
the orbital eccentricity, the libration frequency is $\omega_o=1$.
The amplitude of the libration $\epsilon$ is given by $\epsilon=2e$
for optical librations. For physical librations, obtained for $\Re
\gg 1$, $\epsilon$ has to be measured or estimated, but
are less than the extreme value $2e$ \cite[see the data
in][]{noir2009experimental}. The theoretical analysis
is the same, and the use of formula (\ref{eq:rate}) with
$\tilde{Ra}=0$ (thermal field negligible) and $\omega_o=1$ gives the
LDEI threshold. To obtain a unique stability diagram for all bodies
in Table \ref{table_Io}, we define the quantity
\begin{eqnarray}
Y_2=\left[ \epsilon\ \beta-\frac{4}{17}\ \Lambda \right]\left[
\alpha\ (1-\eta)\ f_{(\eta)} \right]^{-1},
\end{eqnarray}
and use the Ekman number based on the thickness $E_k=E/(1-\eta)^2$.
The threshold for LDEI given by formulas (\ref{eq:rate}) and
(\ref{visccorrect})
 \begin{eqnarray}
 \sigma = \frac{17}{64}\ \epsilon \beta - \alpha\ (1-\eta)\ f_{(\eta)} \sqrt{E_k}-\frac{1}{16}\  \Lambda \geq 0
\end{eqnarray}
is then equivalent to
 \begin{eqnarray}
 Y_2 \geq \frac{64}{17} \sqrt{E_k}.
\end{eqnarray}
This allows us to plot the stability diagram shown in figure
\ref{cebronfig11} in the extreme case of optical libration
($\epsilon=2e$) for a quasi-equilibrium hydrostatic bulge calculated
with formula (\ref{eq:newton2}), corresponding to the optimal case
for LDEI (i.e. the maximum possible libration amplitude and the
maximum possible elliptical deformation). In the following, we
discuss the stability versus the LDEI of the Galilean moons, Titan,
and three Super-Earths. Because the tidal bulge and the libration
amplitudes of these bodies are not yet known \cite[see the
discussion of][]{goldreich2010elastic}, we present the results in
figures (\ref{fig:Io})-(\ref{fig:superearth}) on diagrams in the
$(\beta,\epsilon)$ plane, taking the full range of
variability of $\beta$ and $\epsilon$ into account. Therefore, the upper right
hand corner will correspond to the optimal case for the LDEI: the
libration of a purely deformable body i.e optical libration with
$\epsilon=2e$ and an hydrostatic bulge. The lower left hand corner
corresponds to the libration of a rigid body (physical librations),
associated with the weak diurnal tides. In the same way, the lower
right hand corner corresponds also to physical librations, but with a
hydrostatic bulge. Finally, the upper left hand corner corresponds to the
libration of a purely deformable body (optical libration) associated
to the small diurnal tides amplitude. The relevant physical
configurations for each body depend on their compositions so is specifically discussed in the following for each of them.

First, we consider Io with the values used in the studies of
\cite{kerswell1998tidal} and \cite{herreman2009effects}, i.e. a
static bulge of ellipticity $\beta=0.006$ and a libration amplitude
assumed to be $\epsilon=2e$ (see Table \ref{table_Io}). As already
found by these authors, Io is unstable with a good level
of confidence (typical growth time of 63 years). However, this is an
optimal unrealistic case, because the ellipticity used is due to a static
bulge ($\Re \gg 1$), and the libration amplitude is taken as equal to
$2e$, as in the deformable case ($\Re \ll 1$). Due to its silicate
mantle, the core of Io is expected to be in the limit $\Re \gg 1$,
and consequently the ellipticity to consider is indeed $\beta=0.006$
but the libration amplitude is instead $\epsilon \approx 1.3 \cdot
10^{-4}$ (see Table \ref{table_Io}), which is $63$ times smaller
than $2e$. With these more realistic values and condidering the presence of Jupiter magnetic field, Io is expected to be stable, unlike what was expected. However, to obtain a better
view of the stability in Io, figure \ref{fig:Io} gives the typical
growth time of the instability for different ellipticities and
libration amplitudes, ranging between the limit cases $\Re \gg 1$
and $\Re \ll 1$, with $\epsilon=2e$ and the diurnal tidal
ellipticity $3e \beta \approx 7 \cdot 10^{-5}$, corresponding to a
diurnal tide amplitude of $130$ m. Future accurate measurements of
the tidal amplitude at the core-mantle boundary and of the libration
amplitude should confirm our prediction.

In the optimal case for instability, figure \ref{cebronfig11} shows
that the liquid core of Europa is unstable, as already suggested by
\cite{kerswell1998tidal}, even when taking the Joule
dissipation due to the presence of Jupiter magnetic field into account. The typical growth time of the instability to be around $400$ years, and the associated dissipation to be on the order of $P
\sim 10^{10}\ W$. This corresponds to  $Q^* \sim 10^7$, and is two orders of magnitude below the conservative
estimation of $3 \cdot 10^{12}\ W$ for the tidal heating rate on
Europa \cite[][]{o2002melt}. In reality, the silicate
mantle of Europa should behave more rigidly. Since the libration
amplitude of the mantle and the amplitude of the tidal distortion at
the core-mantle boundary are not known yet, all intermediate
behaviors are explored in figure \ref{fig:europa}a. We conclude that
high libration amplitude and/or rather large elliptical distortion
are needed for Europa's core to be unstable; nevertheless, the parameter range for instability is rather wide in figure \ref{fig:europa}a and seems to be reachable: we thus expect the core
of Europa to be unstable. Results for Europa subsurface ocean are
shown in figure \ref{fig:europa}b. The elastic behavior of the icy
crust above the subsurface ocean is expected to behave in the
deformable limit ($\Re \ll 1$), which correspond to parameters close
to the upper right hand corner. As also shown in figure
\ref{cebronfig11}, Europa's ocean is therefore unstable.

Concerning the two last Galilean moons, the core of Ganymede and the
subsurface oceans of Callisto and Ganymede are found to be stable in
the optimal case (fig. \ref{cebronfig11}). An LDEI is improbable
today. Figure \ref{cebronfig11} also shows that the subsurface ocean of Titan is
probably stable, because even in the optimal case for the LDEI, it
remains in the vicinity of the threshold.

\subsubsection{Super-Earths}

\begin{figure}                    % Chaque figure doit avoir pour nom nomfig1.eps,
  \begin{center}
    \begin{tabular}{ccc}
      \setlength{\epsfysize}{5.0cm}
      \subfigure[]{\epsfbox{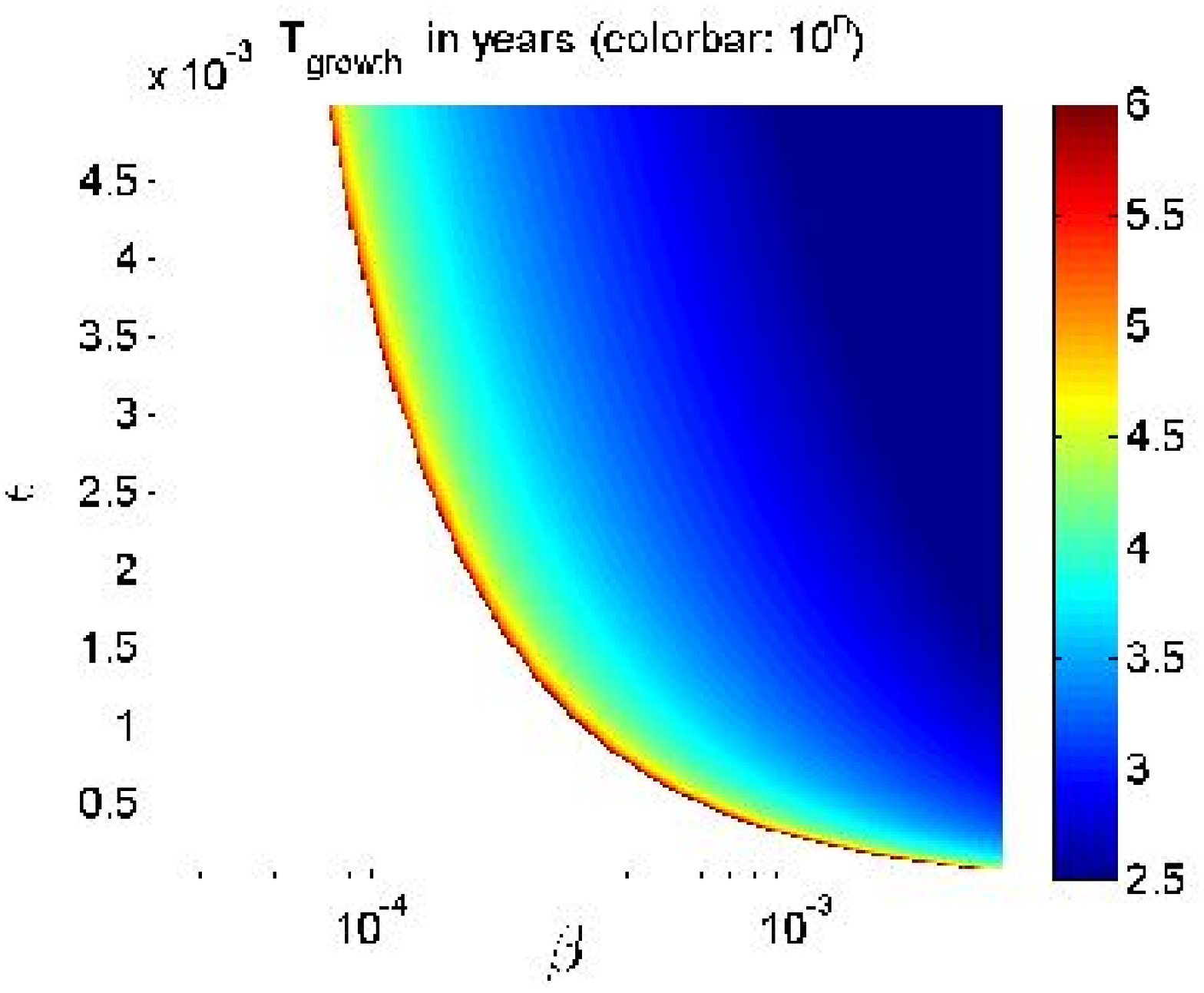}} \\
      \setlength{\epsfysize}{5.0cm}
      \subfigure[]{\epsfbox{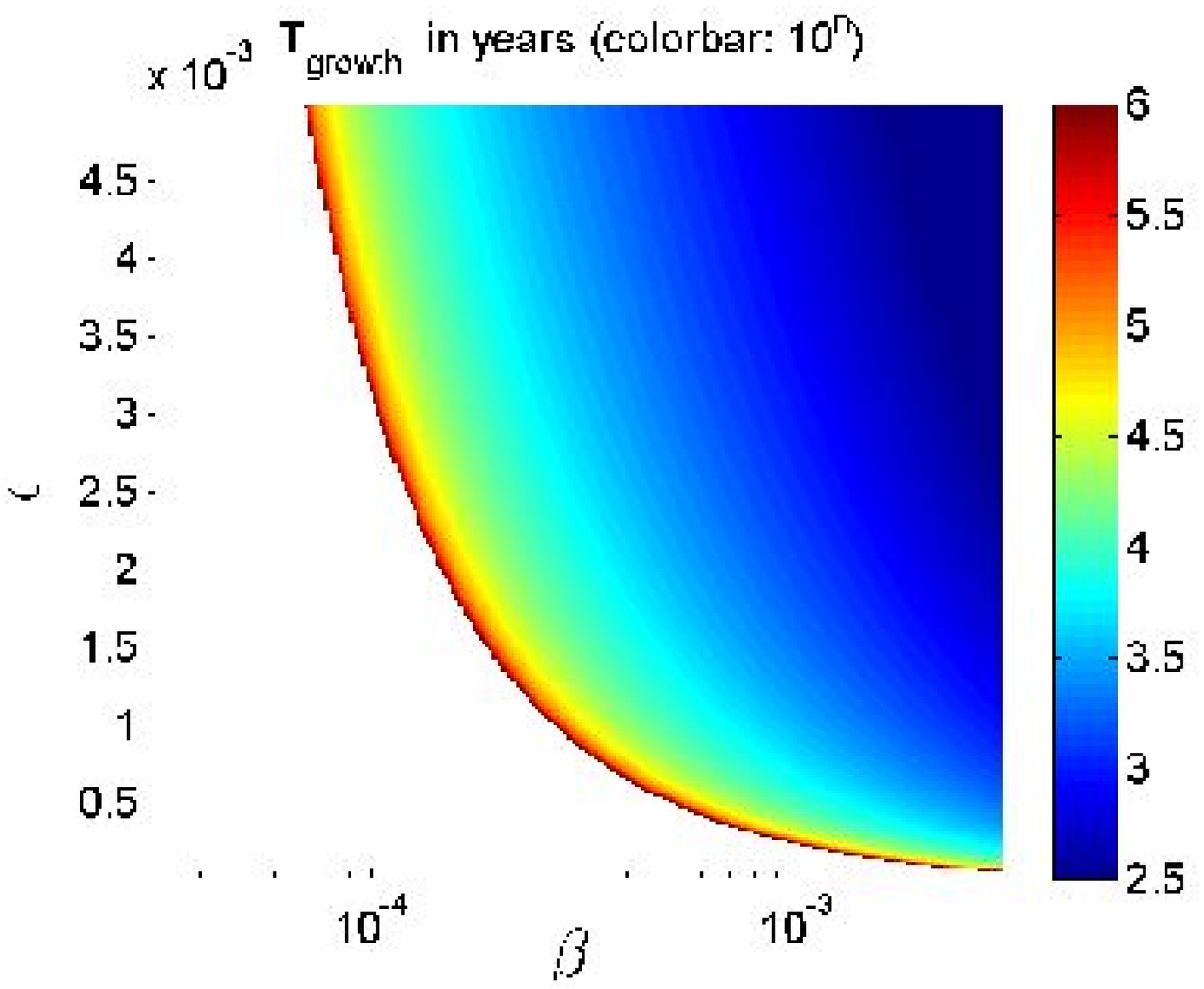}} \\
      \setlength{\epsfysize}{5.0cm}
      \subfigure[]{\epsfbox{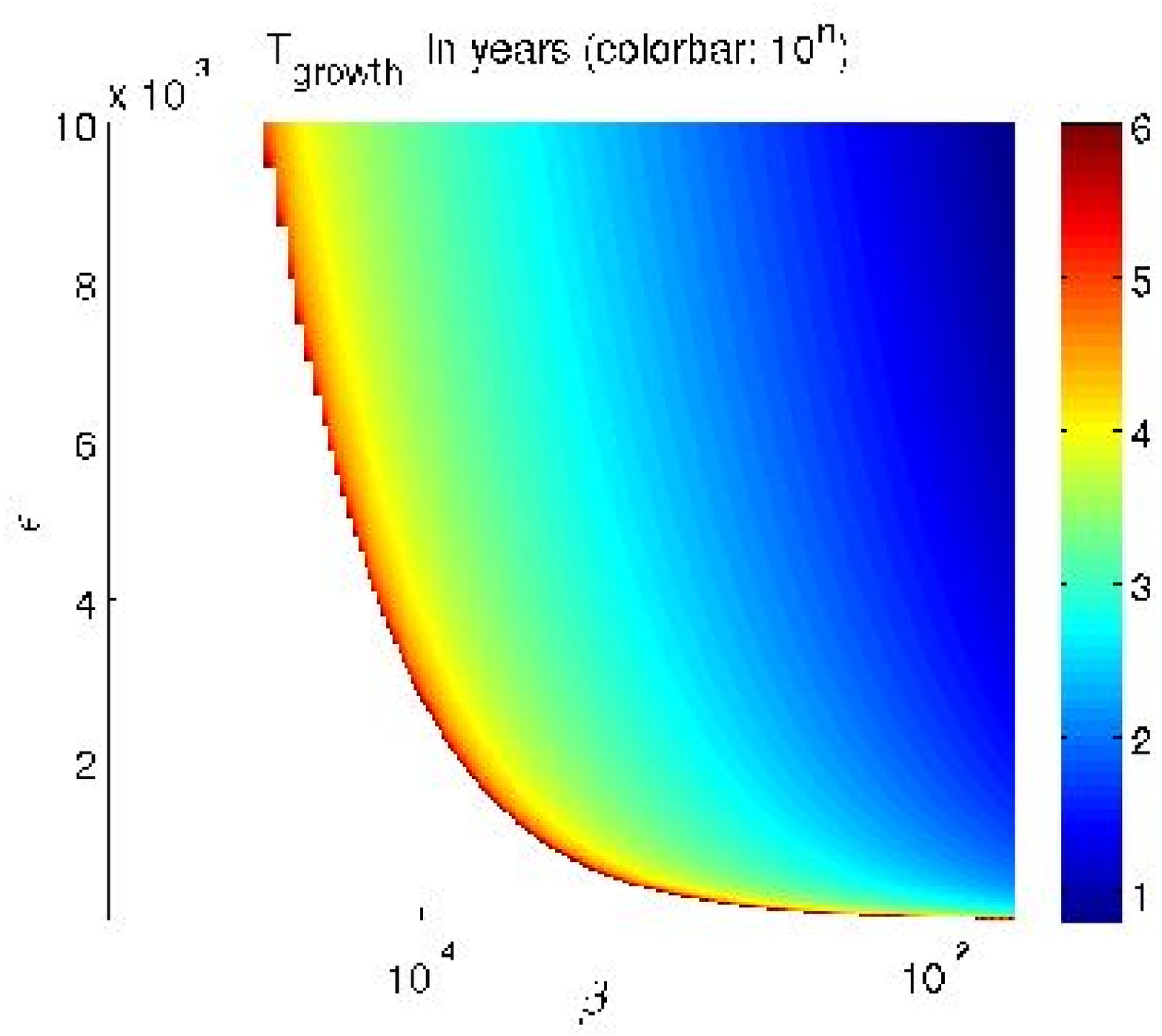}}
    \end{tabular}
    \caption{Same as figure \ref{fig:Io} but for CoroT-7b in figure (a), GJ 1214b in figure (b), and 55 CnC e in figure (c).}
    \label{fig:superearth}             %
  \end{center}
\end{figure}

The recent discovery of extrasolar telluric planets gives typical
examples of synchronized planets in close orbit around their host
stars. This particular astrophysical configuration should lead to a
vigorous LDEI in their possible liquid cores. In this section, we
consider three Super-Earths, expected to be telluric: 55 CnC e,
CoRoT-7b, and GJ 1214b, at $D=0.0156\ \textrm{A.U}$,
$D=0.0172\ \textrm{A.U}$, and $D=0.0143\ \textrm{A.U}$ from their
host star respectively. The data used here are from \cite{winn2011super} for 55
CnC e, \cite{valencia2010composition} for CoroT-7b, and
\cite{charbonneau2009super} for GJ 1214b. They are given in Table
\ref{table_earth}. For CoRoT-7b, the work of \cite{leger2011extreme}
predicts a core composed of liquid metal, representing $11 \%$ of
the total planetary volume, as for the actual Earth. The presence of
a liquid core in Super-Earths is still not clear \cite[e.g.][for
CoroT-7b]{wagner2011physical}, but we can reasonably assume that a
planetary liquid core could occupy about one third of the planet
radius, which corresponds to $4\ \%$ of the total planetary volume.
Because of the proximity of the parent star, these extrasolar
planets are expected to be synchronized. The actual orbital
eccentricities of CoRoT-7b and GJ 1214b are not known. If they are
fully circularized and synchronized, no elliptical instability can
grow. In contrast, if a small libration exists, previous
stability formula can be used. We assume here an orbital
eccentricity of $e=0.001$, well beyond the detection limit. Figure
\ref{cebronfig11} shows that in the optimal case these three
Super-Earths cores are clearly unstable with a good level of
confidence, which means that the LDEI is probably present in their
liquid layers. One can see that this result is not very sensitive
to the hypothesis on the size of the considered liquid core: for
instance, with the values of Table \ref{table_earth}, CoroT-7b
becomes stable for a liquid core aspect ratio $R_2/R$ below $1\
\%$ (for the optimal case i.e. optical librations and equilibrium
tides). In these estimates, we use the hydrostatic tidal
deformation, which underestimates the real tidal deformation. Also, higher orbital eccentricity would lead to more unstable configurations. Figure \ref{fig:superearth} shows the influence of
these uncertainties, as well as the effect of smaller libration
amplitudes and tidal deformations. The proximity of the known Super-Earths with their host stars leads to strong tidal deformations, and the LDEI is then able to grow from very small
libration amplitudes, as shown in figure \ref{fig:superearth} (the true tidal deformation, larger than the hydrostatic value, leads to an LDEI for a large range of libration amplitudes). Same
conclusions are obtained for Kepler-10b
\cite[][]{batalha2011kepler}, assuming the same hypothesis. To conclude, the presence of the LDEI in their liquid cores is very probable, whatever the uncertainty ranges.

\begin{table}
\caption{Physical and orbital characteristics used for the stability
calculations in exoplanets.}

\label{table_earth}
 \begin{tabular}{@{}lcccccc}
      & CoRoT-7b & GJ 1214b & 55 CnC e  \\
 $M$ (in Earth's mass) & $4.8$ & $6.55$ & $8.57$    \\
 $R$ (in Earth's radius) & $1.68$ & $2.678$ & $1.63$ \\
 $T_{orb}\  (d)$ & $0.854$ & $1.58$ & $0.7365$ \\
 $R_2/R$ & $1/3^a$ & $1/3^b$ & $1/3^c$  \\
 $e$  & $0.001^c$ & $0.001^c$ & $0.057^d$ \\
 $\eta$ & $0$ & $0$ & $0$\\
 $E$ ($\times 10^{16}$) & $9.4$ & $6.8$ & $8.6$\\
 $\beta$ ($\times 10^{3}$) & $7$ & $6$ &$5$ \\
$\sigma\ (yr^{-1})$ & $0.01$ & $0.005$ & $0.45$ \\
 \end{tabular}\\

We take as typical value $\nu=10^{-6}\
m^2\ s^{-1}$, consistent with a Fe/Fe-S composition. We use the
formula (\ref{eq:newton}) to estimate the tidal bulge ellipticity.\\

$^a$ Coherent with \cite{leger2011extreme}.\\
$^b$ Coherent with the values from \cite{nettelmann2011thermal}. \\
$^c$ Assumed hypothesis.\\
$^d$ \cite{winn2011super}.

\end{table}

\section{Conclusion and discussion}

In conclusion, we have investigated theoretically the elliptical instability in telluric celestial bodies. New
analytical results were determined to fill the gap between
previous studies and astrophysical applications. In particular, we have derived generic formulas for the growth rate of the elliptical instability
driven respectively by tides in non-synchronized bodies (TDEI) and
libration in synchronized ones (LDEI), in the
presence of imposed magnetic and thermal base fields. It was shown that an elliptical instability is strongly expected in the core of Europa, 55 CnC e, CoRoT-7b, and GJ 1214b, as well as in the
subsurface ocean of Europa. Those results are valid for the present
state of the considered bodies and do not preclude any elliptical instability in the past. For instance, the Early Earth core was clearly unstable, because of the larger gravitational
distortions when the Moon was closer.

One can wonder about the signatures and consequences of such
an instability on the planetary dynamics. A first consequence would
be on the orbital evolution and synchronization process: indeed, the
elliptical instability generates three-dimensional turbulent flows
with cycles of growth, saturation, fluctuations, and relaminarization
\cite[e.g.][]{le2010tidal}. Timescales involved range typically
between the spin period and the growth time of the instability.
Dissipation rates on the planetary scale, and consequently the
orbital evolution, may then follow the same variations, with periods
of rapid evolution when an elliptical instability is present,
followed by more quiescent periods, for instance when the forbidden
zone is reached. This increased dissipation should accelerate the
synchronization process, as described in \cite{le2010tidal}, and
this range of timescales should appear in the evolution of the spin
rotation rate.

The second consequence would be on heat flux variations at the
planetary surface. Indeed, as shown in \cite{cebron2010tidal,
lavorel2010experimental}, flows driven by elliptical instability are
very efficient in transporting heat by advection. As a result,
subadiabatic cores should not be regarded as thermal blankets in the presence of the elliptical instability. The usual Nusselt number $Nu=Q_{tot}/Q_{diff}$, the ratio between the total outward heat flux $Q_{tot}$ and the purely diffusive outward heat flux $Q_{diff}$, which is associated to this heat
advection, is given by the following scaling law, verified both
experimentally \cite[][]{lavorel2010experimental} and numerically
\cite[][]{cebron2010tidal}:
\begin{eqnarray}
Nu=\frac{0.01}{\sqrt{E}}.
\end{eqnarray}
This leads to a total outward heat flux advected by the elliptical
instability about $Nu \approx 3 \cdot 10^4$ times greater than a
purely diffusive outward heat flux. Besides, in the presence of
natural thermal convection, the superimposition of chaotic
elliptically driven flows would induce large-scale variations in the
same amplitude.

Finally, internal
flows driven by elliptical instability are directly responsible for
magnetic field generation. The question of whether LDEI and
TDEI are dynamos-capable is still open and remains out of the reach of
the currently available numerical capacity, but elliptically driven flows induce a magnetic field from an existing background one. To estimate a typical amplitude of such an
induced field, we can use the results of our WKB approach. This
shows that the dimensionless induced magnetic field inside the core and the
instability velocity $\mathbf{u_0}$ are systematically related by
\begin{equation}
 \mathbf{B}= \textrm{i}\ \frac{Rm\ k_{x_3}}{k^2}\ \mathbf{u_0}, \label{MHD_eq}
\end{equation}
where $k$ and $k_{x_3}$ are the norm and the axial
component, respectively, of the wave vector of the excited mode of the elliptical
instability.  This generic expression shows that the induced
magnetic field due to the elliptical instability is systematically
proportional to and in quadrature with the velocity field due to the
instability. For the TDEI, $k_{x_3}=k/2$, and for the LDEI, $k_{x_3}=\omega_o\ k/4$ (see respectively appendices \ref{WKB1} and \ref{WKB2}). Then, assuming that at saturation, the
typical flow induced by instability is commensurate with the
differential rotation between the fluid and the elliptical
distortion (see Table \ref{table_intro}), we estimate the surface
field by
\begin{eqnarray}
b_{surf}= \frac{Rm\ k_{x_3}}{k^2}\ \frac{\bigtriangleup
\Omega}{\Omega} \left( \frac{R_2}{R} \right)^3,
\end{eqnarray}
where $R$ is the planet radius. Starting from the Jovian magnetic
field component along the rotation axis, the LDEI in Europa
subsurface ocean is capable of inducing surface variations of up to
$\sim 0.1\ \%$ of the ambient field (reached for $k=2 \pi/(1-\eta)$
and optical librations, i.e. $\epsilon=0.0188$), and LDEI in its
core up to $\sim 100\ \%$ of the ambient field at the surface
(reached for $k=2 \pi/(1-\eta)$ and physical librations, i.e.
$\epsilon=2 \cdot 10^{-4}$).  Considering Galileo's E4 flyby of
Europa \cite[see][]{zimmer2000subsurface,kabin1999europa}, the
background z-component of the magnetic field is modified from $410$
nT to $380$nT at a distance of Europa about $1.5$ Europa radius.
This modification of $30$ nT is on the same order of magnitude as the
possibly LDEI induced magnetic field from the core. Internal sources
should thus be considered in addition to plasma currents
\cite[][]{kabin1999europa} for interpretating Europa's magnetic
signal.

To finish with, one should notice that the TDEI and LDEI studied here for telluric planets, can also affect the giant gaseous planets of our solar system \cite[][]{wicht2010theory}, as well as extrasolar gaseous
planets such as the hot-Jupiters, whose dramatic tidal deformations
should excite vigorous elliptical instabilities in the planetary
atmospheres, but also in their host stars
\cite[][]{rieutord2003evolution,ou2007further,cebron2010ohp}.

\begin{acknowledgements}
      The authors acknowledge S. Le Diz\`es for his valuable help on the stability analysis; J. Noir and A. Sauret for fruitful discussions on libration; J. Besserer and G. Tobie for fruitful discussions and values on tidal heating in the Galilean moons and Titan.
\end{acknowledgements}

\appendix

\section{Stability analysis for non-synchronized systems} \label{WKB1}
Solving (\ref{eq:streamlines}), we find the trajectories
\begin{eqnarray}
\mathbf{x}= r\
\left [
   \begin{array}{ccc}
      \sqrt{1+\beta}\ \cos (t \sqrt{1-\beta^2}) \\
      \sqrt{1-\beta}\ \sin (t \sqrt{1-\beta^2})  \\
       0 \\
          \end{array}
   \right ] \label{eq:streamline1}
\end{eqnarray}
\noindent  where without loss of generality, the origin of time has
been defined such that the initial position along the trajectory is
$\mathbf{x}_{(t=0)}=[x_1,x_2,x_3]=[r\ \sqrt{1+\beta}\ , 0, 0]$ (the results
of the WKB analysis do not depend on this chosen initial position).

The solution to equation (\ref{eq:wave}) at order $1$ in $\beta$ along
a streamline (\ref{eq:streamline1}) writes as
\begin{eqnarray}
\mathbf{k}=
\left [
   \begin{array}{ccc}
      k_{x_1}\\
      k_{x_2} \\
      k_{x_3} \\
   \end{array}
   \right ]=
\left [
   \begin{array}{ccc}
      {k_{x_1}}_o\ \cos t - {k_{x_2}}_o\ \sin t +\beta \  {k_{x_2}}_o\ \sin t \\
      {k_{x_1}}_o\ \sin t + {k_{x_2}}_o\ \cos t +\beta \  {k_{x_1}}_o\ \sin t \\
      k_0 \cos{a} \\
   \end{array}
   \right ] \label{stream_k}
\end{eqnarray}
where ${k_{x_1}}_o$, ${k_{x_2}}_o$, and $k_0$ are constant, and $a$ is
the angle between the wave vector and the rotation axis. We define
the phase $\phi$ of the wave vector by writing ${k_{x_1}}_o=k_0
\sin{a} \ \cos \phi $, ${k_{x_2}}_o=k_0 \sin{a} \ \sin \phi $.

We are now in a position to solve the system of linearized equations
given in section \ref{validation}. To do so, we use as unknowns the
vertical velocity of the perturbed field  $u_3$ and its vertical
vorticity $W_3=\partial_{x_1} u_2-\partial_{x_2}
u_1=\I(k_{x_1}u_2-k_{x_2}u_1)$, as well as the vertical component
$b_3$ of the perturbed magnetic field and the corresponding magnetic
vertical vorticity $C_3=\I (k_{x_1}b_2-k_{x_2}b_1)$. The resolution
is then straightforward \cite[see][]{herreman2009effects}.

At order $0$ in $\beta$, the system reduces to an harmonic equation for $u_3$,
giving a dispersion relation with a pulsation $f=2\ (1+\Omega^G)
\cos a $, with the quantity
$\Omega^G=\Omega_{orb}/(\Omega_{spin}-\Omega_{orb})$ already used by
\cite{kerswell2002elliptical,le2010tidal}. Solvability conditions
imply non-trivial solutions only if $f=1$, which gives the resonance
condition $\cos a=1/(2\ (1+\Omega^G)) \in [-1, 1]$. This means that
the instability cannot grow when $\Omega/\Omega_{orb} \in [-1;1/3]$,
which is the so-called forbidden zone. Outside this band, the growth
rate is determined by the nullity of the determinant of the
solvability condition system. It is then maximized over all values
of wave vector phase $\phi$. The maximum is obtained for
$\phi=\pi/4$ and the inviscid growth rate writes as
\begin{eqnarray}
\sigma_{inv} &=& \frac{(2\Omega^G+3)^2}{16\ |1+\Omega^G|^3}\  \sqrt{\beta^2-4\ \zeta^2}\ -\frac{\Lambda}{4\ |1+\Omega^G|^3\ (1+Rm^2 k^{-4})} \label{rota}  \nonumber\\
& & {}- \frac{k^2\ \tilde{Ra}\ r\ \partial_r \theta}{8\ (1+k^4\ E^2/Pr^2)\ |1+\Omega^G|^3} , 
\end{eqnarray}
with
\begin{eqnarray}
\zeta=\frac{(k^4+Rm^2)\ \tilde{Ra}\ r\ \partial_r \theta -2\ k^2\ Rm\ \Lambda\ (1+k^4\ E^2/Pr^2)}{(2\Omega^G+3)^2\ (1+k^4\ E^2/Pr^2)\ (k^4+Rm^2)},
\end{eqnarray}
where $r$ and $\partial_r \Theta$ are the radius and
the dimensionless radial gradient of temperature base field on the
considered streamline, respectively. In the absence of viscous boundary damping
(discussed in section \ref{damping_term}), the inviscid growth rate
is a correct approximation of the viscous growth rate when the
viscous diffusion term $-k^2 E$ in equation (\ref{eq:NS}) is
negligible, i.e. for perturbations of wavelength greater than the
Ekman thickness $\sqrt{E}$. Equation (\ref{rota}) takes the thermal and magnetic diffusions into
account. In a typical liquid
core, the thermal Prandtl number is about $O(0.1 - 1)$. Neglecting
the viscous diffusion term also leads to neglecting the thermal
diffusion term $-k^2 E/Pr$ of equation (\ref{eq:WKB_temp}). But
since the magnetic Prandtl number is about $Pm \approx 10^{-6}$
\cite[e.g.][]{christensen2006scaling}, the magnetic diffusion term
$-k^2/Rm=-k^2 E/Pm$ of equation (\ref{eq:WKB_magn}) appears as the
dominant diffusion mechanism. It can nevertheless be neglected within
the limit of large wavenumbers, where equation (\ref{rota}) gives
equation (\ref{eq:general}).

\section{Stability analysis for synchronized systems} \label{WKB2}

We consider that the differential rotation between the fluid and the
elliptical distortion writes as $1-\gamma_{(t)}=\epsilon \cos (\omega_o
t + q)$. The phase q is introduced here because we define
the origin of time such that the initial position of the considered
trajectory is $x_2=0$. Equation (\ref{eq:streamlines}) can be solved analytically
to find the trajectories. In the particular case $q = 0$, the
solution can be written in the following compact form:
\begin{eqnarray}
\mathbf{x}= s_g\ r\
\left [
   \begin{array}{ccc}
       \displaystyle  \sqrt{1+\beta}\ \cos \left (\frac{\epsilon}{\omega_o}\ \sin (\omega_o t)\  \sqrt{1-\beta^2} \right) \\[3 mm]
       \displaystyle  \sqrt{1-\beta}\  \sin \left (\frac{\epsilon}{\omega_o}\ \sin (\omega_o t)\ \sqrt{1-\beta^2} \right)  \\
       0 \\
   \end{array}
   \right ] \label{streamq0_2}
\end{eqnarray}
with $s_g=\sgn (\cos (\omega_o t))$. In the general case, the wave vector associated  to
this flow writes as
\begin{eqnarray}
\mathbf{k}=
\left [
   \begin{array}{ccc}
      k_{x_1}  \\
      k_{x_2}  \\
      k_{x_3}  \\
   \end{array}
   \right ]=
\left [
   \begin{array}{ccc}
     \displaystyle  {k_{x_1}}_o\ -\frac{{k_{x_2}}_o}{M} \left ( \sin (\omega_o t+q)  - \sin q \right) \left (1 - \beta \right) \epsilon  \\[3 mm]
     \displaystyle {k_{x_2}}_o\ +\frac{{k_{x_1}}_o}{M} \left ( \sin (\omega_o t+q)  - \sin q \right) \left (1 + \beta \right) \epsilon   \\
      k_0 \cos{a} \\
   \end{array}
   \right ] \label{stream_k2}
\end{eqnarray}
where ${k_{x_1}}_o$, ${k_{x_2}}_o$, and $k_0$ are constant and $a$ is
the angle between the wave vector and the rotation axis. We define
again the phase $\phi$ by ${k_{x_1}}_o=k_o\ \sin{a} \ \cos \phi $,
${k_{x_2}}_o=k_o\ \sin{a} \ \sin \phi $.

At leading order in $\epsilon \beta$, the dispersion relation gives $f=2\
\cos a$, and the solvability conditions system admits non-trivial
solutions for $f=\omega_o/2$. Consequently, the authorized band is
given by $\cos a=\omega_o/4 \in [-1, 1]$ i.e. $|\omega_o| \leq 4 $.
The growth rate is determined at order 1 in $\epsilon \beta$ and
must be maximized above the phases $q$ and $\phi$. The maximum is
reached for $q=0$, $\phi=\pi /4$ and gives in the absence of thermal
and magnetic fields
\begin{eqnarray}
\frac{\sigma_{inv}}{\epsilon \beta}= \frac{16+\omega_o^2}{64}.
\end{eqnarray}

\cite{kerswell1998tidal} have performed a global approach to the same
instability, explicitly considering inertial waves coupling in a
spheroidal geometry. In the absence of magnetic and thermal fields,
they found a maximum inviscid growth rate $\sigma_{inv}/(\epsilon\
\beta)=25/128$, very close to our value $\sigma_{inv}/(\epsilon\
\beta)=17/64$. The small difference between the two values is due to
the influence of the spheroidal geometry considered in
\cite{kerswell1998tidal}, leading to more restrictive conditions for
destabilization than our local analysis. Similarly, for purely
hydrodynamic flow with a stationary deformation as studied in
section \ref{sec:background}, our analysis gives the inviscid growth
rate $\sigma_{inv}/\beta=9/16$, whereas a global analysis with
inertial waves of a spheroid leads to the slightly lower value
$\sigma_{inv}/\beta=1/2$
 \cite[see][]{lacaze2004elliptical}. Taking the uncertainties on
the different parameters for planetary application into account, these small
differences can be disregarded and the local approach can be used confidently, which presents the strong advantage of providing an explicit
formula for the growth rate.

Taking a buoyancy of order $\epsilon \beta$ into account, as well as the induction equation and a Lorentz force on the order of $\epsilon \beta$ in the presence of an imposed vertical magnetic field
$B_0$, we obtain the growth rate
\begin{eqnarray}
\sigma_{inv} &=&  \frac{16+\omega_o^2}{64}\ \sqrt{(\epsilon \beta)^2-4\ \omega_o^2\ {\zeta_2}^2} \label{rota2}  \nonumber\\
& & {} -\frac{\omega_o^2 \Lambda}{16\ (1+k^{-4} \omega_o^2 Rm^2/4)} - \frac{(16+\omega_o^2)\ k^2\ \tilde{Ra}\ r\ \partial_r \theta}{16\ (\omega_o^2+4\ k^4\ E^2/Pr^2)}
\end{eqnarray}
with
\begin{eqnarray}
\zeta_2=\frac{F}{(16+\omega_o^2)\ (4\ k^4+\omega_o^2\ Rm^2)\ (\omega_o^2+4\ k^4\ E^2/Pr^2)}
\end{eqnarray}
where $F=(64+\omega_o^2 (4 k^4+Rm^2 (16+\omega_o^2)))  \tilde{Ra}\ r\ \partial_r \theta -4 k^2 \omega_o^2 Rm \Lambda (\omega_o^2+4 k^4 E^2/Pr^2)$, and where $r$ and $\partial_r \theta$ are respectively the radius and the dimensionless temperature radial gradient of the considered streamline. As discussed in appendix \ref{WKB1}, in astrophysical applications, the thermal diffusion can be neglected ($E/Pr=0$). In this case, equation (\ref{rota2}) gives equation (\ref{eq:rate}) in the limit of large wavenumbers.

\section{Is the diurnal tide stabilizing or destabilizing for the elliptical instability?} \label{sec:diurnal}

In this work, the periodic forcing due to diurnal tides has been
neglected for synchronized bodies. It is thus legitimate to
calculate its influence on the LDEI growth. To answer this question,
we consider the simplest but severe case of a body with a non
rotating (i.e. $\gamma = 0$) diurnal tide of amplitude
$\beta_{(t)}=\beta_1 \cos (M t+q)$, where $\beta_1=3e \beta$ and there is no
global rotation. Then, the base flow (\ref{baseflow}) reduces to
\begin{eqnarray}
\mathbf{U} =  [-(1+\beta_1 \cos (M t &+ &q))\ x_2\ \mathbf{e_{x_1}}\nonumber\\ [3mm]
& & +(1-\beta_1 \cos (M t+q))\ x_1\ \mathbf{e_{x_2}}]
\end{eqnarray}
Once again, the phase $q$ is introduced here because we fix the
phase of streamlines. The streamlines are not known
analytically. The dispersion relation gives the pulsation $f=2 \cos
a$, and the solvability conditions give resonances for $f=(2+M)/2$
and $f=(2-M)/2$. The authorized band is thus $|M| \leq 6 $. In the
limit of low $M$, the maximal growth rate is obtained with
$f=(2-M)/2$ for $\phi=-\pi/4$ and $q=0$:
\begin{eqnarray}
\sigma_{inv}= \left[\frac{9}{16}+\frac{81}{64}\ M+\left (\frac{1}{8}-\frac{15 \pi^2}{32}\right) M^2 \right] \beta_1  \label{eq:diurnal}
\end{eqnarray}
at the order $O(\Lambda^2)+O(\beta_1\ M^3)+O({\beta_1}^2)$. This expression agrees with equation (\ref{eq:general}) for $\Lambda=0$ and $\tilde{Ra}=0$, in the
limit $M=0$. The expression (\ref{eq:diurnal}) shows that slow
oscillations of the amplitude of the tidal bulge are not inhibiting
for the elliptical instability. On the contrary, the growth rate is
enhanced compared to the case of constant amplitude for low $M$,
which means that the diurnal tide would be destabilizing in this
case. In the case of planetary interest with $M=1$, the maximum
growth rate writes as
\begin{eqnarray}
\sigma_{inv}=\frac{25}{128}\ \beta_1,
\end{eqnarray}
which again shows that the diurnal tide can drive an elliptical
instability. This effect will thus be superimposed on the TDEI and LDEI
mechanisms already studied, but with a slower growth rate, since $\beta_1 \ll \beta$.

\section{Resonances of gravito-inertial waves} \label{GI}

To clarify the influence of a thermal field with a buoyancy
force that is on the order of zero in $\beta$, we consider the generic case
with the elliptical gravitational iso-potentials of ellipticity $n\
\beta$ and elliptical base-field isotherms of ellipticity $m\
\beta$, where $n$ and $m$ are arbitrary constants. This generic
notation is needed to deal with all cases, as studied for instance
in \cite{le2006thermo}, \cite{lavorel2010experimental} and
\cite{cebron2010tidal}. In focusing on dynamic tides in a
non-synchronized system, one would expect the isotherms to follow
the streamlines (because of the small thermal diffusion
coefficient), as well as the iso-potentials. Hence, $n=m=1$. In contrast, looking at the elliptical instability in a subsurface ocean underlain by a rigid mantle, one would expect the
iso-potentials to remain quasi-circular ($n=0$). Besides this, in the
presence of a static bulge, one would expect the system to naturally
return to a configuration with $m=0$ by the generation of baroclinic
motions. All situations with $0 \leq m,n \leq 1$ are possible, and
one can even imagine other azimuthal periodicities, for instance those due
to local variations of temperature.

In the case of the TDEI, the WKB approach including zeroth-order
buoyancy forces in $\beta$ is tractable. Unlike the cases
studied in appendices \ref{WKB1} and \ref{WKB2}, the forbidden
band where the instability does not exist is now modified by the
thermal field and is given by $f_0 \leq 1$, where $f_0=\sqrt{4\
|1+\Omega^G|^2+\tilde{Ra}\ r\ \partial_r \Theta}$. Outside this
band, the inviscid growth rate is given, in the limit of large
wavenumbers, by:
\begin{eqnarray}
\frac{\sigma_{inv}}{\beta} &=& \frac{(2 \Omega^G+3)^2+ [1+2 (1+\Omega^G)
(m-n)-n]\ \tilde{Ra}\ r \partial_r \Theta}{16\ |1+\Omega^G|^3+4\
\tilde{Ra}\ |1+\Omega^G|\ r\ \partial_r \Theta} \label{eq:TDEIorder00} \\ \nonumber & & -
\frac{\Lambda}{4\ |1+\Omega^G|^3}\ \frac{1}{\beta}.
\end{eqnarray}
We can compare the role of the temperature field in
(\ref{eq:general}) and (\ref{eq:TDEIorder00}). In expression
(\ref{eq:general}), the temperature field acts as a simple
supplementary stabilizing term that corrects the inviscid purely
hydrodynamic growth rate. But in the derivation of equation
(\ref{eq:TDEIorder00}), waves and resonances (as well as the
forbidden band) are modified by buoyancy forces, leading to a
modification of the prefactor of $\beta$. Actually, the elliptical
instability now results from resonances of gravito-inertial waves
and should be called the gravito-elliptical instability
\cite[see][]{le2006thermo,guimbard2010elliptic}. As shown below, the
supplementary resonances associated to gravito-inertial waves allow
the temperature to be destabilizing in certain cases. The same
conclusions can be obtained for the magnetic field when Lorentz
forces are taken into account at zeroth order in $\beta$ in the
limit of ideal magnetohydrodynamic: the elliptical instability then
results from resonances between magneto-inertial waves
\cite[][]{kerswell1993elliptical,kerswell1994tidal}; the forbidden
band is modified by the magnetic field; and as shown by
\cite{lebovitz2004magnetoelliptic}; and the magnetic field can be either
stabilizing or destabilizing depending on the case being considered
\cite[see
also][]{Herreman_PhD,mizerski2009magnetoelliptic,mizerski2011influence}.
Naturally, these conclusions are also valid for the LDEI.

\begin{figure}
  \begin{center}
    \epsfysize=6.0cm
    \leavevmode
    \epsfbox{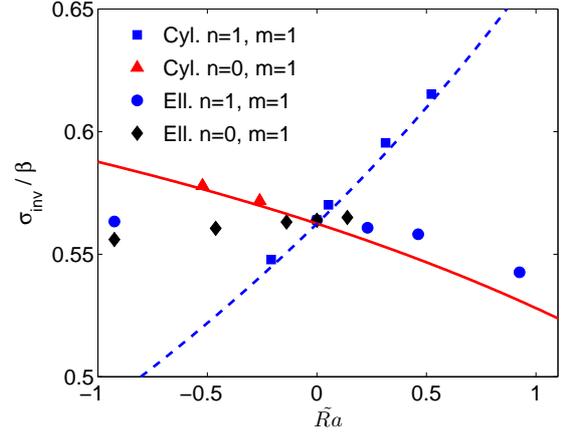}
    \caption{Growth rate of the TDEI for a cylindrical shell  of aspect ratio
    $H/R=2$ with an elliptical cross section ($\eta=0.2,\ E=0.0036,\ \beta=0.47,\ Pr=1)$ and an ellipsoidal shell ($\eta=0.3,\ E=0.0029,\ \beta=0.317,\ Pr=1)$ with a rotation axis of length $c=(a+b)/2$.
    The figure compares the numerical growth rate in the autogravitating case where the gravity is given by the Poisson equation
    for the gravitational potential of a homogeneous fluid \cite[see][for details]{cebron2010tidal}, and the case
    where the gravity is played by a centrifugal force ($n=0$), as in the experiments of \cite{lavorel2010experimental}.
    The numerical growth rate is translated vertically in the figure, to match the inviscid growth rate $9/16$ at $\tilde{Ra}=0$,
    which corresponds to a surfacic damping term coefficient $\alpha = 3.24$ for the cylindrical shell (squares and triangles)
    and $\alpha=3.1$ for the ellipsoidal shell (circles and diamonds). Theoretical growth rates are shown by a continuous red line for
    $n=0,\ m=1$ and a dashed blue line for $n=m=1$.}
    \label{cebronfigthermiq}
  \end{center}
\end{figure}

In \cite{le2006thermo} and in the experiments of
\cite{lavorel2010experimental}, the TDEI is studied for a
stationary bulge ($\Omega^G=0$) with circular iso-potentials and
elliptical isotherms ($n=0$, $m=1$), and the considered temperature
profile gives $r\ \partial_r \Theta=-1$. In this particular case, in
the absence of magnetic field, equation (\ref{eq:TDEIorder00})
recovers their result:
\begin{eqnarray}
\sigma_{inv}= \frac{9-3\ \tilde{Ra}}{16-4\ \tilde{Ra}}\ \beta.
\end{eqnarray}
As already noticed by \cite{le2006thermo}, a thermal stable stratification
($\tilde{Ra}<0$) is then destabilizing for the elliptical
instability, but, in constrast, the temperature field stabilizes the instability for $n=m=1$. This high sensitivity of the growth rate of the elliptical instability to the specific
gravitational and thermal fields is confirmed by numerical
simulations. Using the method described in \cite{cebron2010tidal},
we consider the simple case $\Omega^G=0$ and $K=0$, the temperature
field being established by a temperature contrast between the two
boundaries. When $n=m=1$, the growth rate
(\ref{eq:TDEIorder00}) is enhanced when $\tilde{Ra}$ is increased.
As shown in figure \ref{cebronfigthermiq}, this is in perfect
agreement with the numerical simulations in a cylindrical shell. In
the experimental setup of \cite{lavorel2010experimental}, the
gravity is replaced by the centrifugal acceleration, as in
\cite{carrigan1983experimental}, and the associated equipotentials
are circular, i.e. $n=0$ and $m=0$, as shown in figure
\ref{cebronfigthermiq}. In this case an increasing $\tilde{Ra}$
indeed leads to a lower growth rate and the numerical simulations agree with the predicted growth rate.

The conclusions for an autogravitating ellipsoidal shell are more
complex. \cite{cebron2010tidal} find that an increasing $\tilde{Ra}$
leads to a lower growth rate, which contradicts the prediction of the theoretical growth rate. This difference comes from the fact that the thermal stratification propagates the influence of the
boundary inside the bulk: the WKB analysis, based on local
stability, cannot handle this feature. In the spherical geometry, we
can, however, notice that changes induced by $\tilde{Ra}$ are small
for $-1 \leq \tilde{Ra} < 1$ and remain close to the estimates for
an autogravitating cylinder.

Those results clearly illustrate the high sensitivity of the growth
rate of the elliptical instability to the specific gravitational and
thermal fields, as well as to the considered geometry. In planetary
applications, stratification (i.e. $\tilde{Ra}<0$) generally leads
to stabilization, as in the limit of small $\tilde{Ra}$ presented in
the main text; however, in this case, stratification can
only stabilize elliptical instability when $\tilde{Ra}=O(1-10)$, which is never the case.

\bibliographystyle{aa}
\bibliography{aa_telluric}

\end{document}